\documentstyle[11pt,aaspp4]{article}

\begin{document}

\title{Tidally-Triggered Star Formation in Close Pairs of Galaxies}
\author{Elizabeth J. Barton\altaffilmark{1}, Margaret J. Geller, and Scott J. Kenyon}
\affil{Harvard-Smithsonian Center for Astrophysics}
\authoraddr{ebarton@cfa.harvard.edu, Mail Stop 10, 60 Garden St., Cambridge, MA 02138}
\authoraddr{mgeller@cfa.harvard.edu, Mail Stop 19, 60 Garden St., Cambridge, MA 02138}
\authoraddr{skenyon@cfa.harvard.edu, Mail Stop 15, 60 Garden St., Cambridge, MA 02138}
\altaffiltext{1}{present address: National Research Council of Canada, Herzberg Institute of Astrophysics, Dominion Astrophysical Observatory, 5071 W. Saanich Road, RR5,
Victoria, BC, Canada V8X 4M6}

\begin{abstract}

We analyze optical spectra of a sample of 502 galaxies in close 
pairs and n-tuples, separated by $\leq 50 {\rm h}^{-1} {\rm kpc}$.  
We extracted the sample objectively from the CfA2 redshift survey,
without regard to the surroundings of the tight systems; we re-measure
the spectra with longer exposures, to explore the spectral characteristics
of the galaxies.  We use the new spectra to probe the relationship 
between star formation and the dynamics of the systems of galaxies.

The equivalent widths of H$\alpha$ (EW(H$\alpha$))
and other emission lines anti-correlate strongly with 
pair spatial separation ($\Delta D$) and
velocity separation; the anti-correlations do not result from any
large-scale environmental effects that we detect.  
We use the measured EW(H$\alpha$) and the 
starburst models of Leitherer et al.
to estimate the time since the most recent burst of
star formation began for galaxies in our sample.  
In the absence of a large contribution from an old stellar population to
the continuum around H$\alpha$ that correlates with the
orbit parameters, the observed $\Delta D$~--~EW(H$\alpha$)
correlation signifies that starbursts with larger separations
on the sky are, on average, older. 
We also find a population of
galaxies with small to moderate amounts of Balmer absorption. These
galaxies suport our conclusion that the sample includes many
aging bursts of star formation; they
have a narrower distribution of velocity separations, consistent
with a population of orbiting galaxies near apogalacticon.

By matching the dynamical timescale to the 
burst timescale, we show that the data support a simple picture 
in which a close pass initiates a starburst;
EW(H$\alpha$) decreases with time as the pair separation
increases, accounting for the anti-correlation.
Recent n-body/SPH simulations of interacting pairs suggest a
physical basis for the correlation ---
for galaxies with shallow central potentials,
they predict gas infall before the final merger.
This picture leads to a method for measuring the duration and
the initial mass function of interaction-induced starbursts:
our data are compatible with the starburst
models and orbit models in many respects, as long as the starburst lasts longer
than $\sim$10$^{8}$ years and the delay between the close pass and
the initiation of the starburst is less than a few$\times 10^7$ years.
If there is no large contribution from an old stellar population to
the continuum around H$\alpha$,
the Miller-Scalo and cutoff (M$ \leq 30\ $M$_{\sun}$) Salpeter initial mass
functions fit the data much better than a standard Salpeter IMF.

\end{abstract}

\keywords{Galaxies: interactions --- galaxies: starburst --- stars: formation ---
stars: luminosity function, mass function}

\section{Introduction}

Constraints on both the merger rate at the current epoch and the
orbits of pairs of galaxies are important cosmological indicators.
Furthermore, cosmological models of galaxy formation depend on 
knowledge of the properties of interacting galaxies.
N-body simulations of the formation and evolution of structure in the
Universe incorporate semi-analytic models of star-formation, merging,
and feedback into the interstellar medium (e.g., Kauffmann et al. 1999;
Somerville et al. 1998a, 1998b).  
A recent model in which galaxy mergers trigger 
intense star formation (Somerville et al. 1998a, 1998b),  
is consistent with measurements of star formation at high
redshift (Hughes et al. 1998).  In this
model, mergers are critical: the ``Lyman break'' 
galaxies (Steidel et al. 1996), at $z \geq 2.8$, 
are moderate-mass galaxies visible due to intense, short-lived 
bursts of star formation triggered by galaxy mergers.
Generally, the semi-analytic merger models are based on 
high-resolution simulations.  
Here, we work toward establishing an observational foundation
for models of galaxy evolution.

Dramatic morphological, kinematic and spectroscopic
signatures of interaction, including tidal tails and streams, distorted
kinematics, extreme infrared emission, 
and starbursts, indicate the effects of interaction in 
individual cases.  
Numerical models reproduce these features, further supporting the 
interpretation that they are tidally-triggered (e.g., Toomre \& Toomre 1972;
more recently, Mihos \& Hernquist 1996).  However, not all 
interacting systems will show these effects.  For example,  
Toomre \& Toomre (1972) showed that
tidal tails and streams are resonance effects which are largely
absent from retrograde encounters.  
Sensitivity of these features to orbit parameters and to the
mass distribution in the galaxies (Dubinski, Mihos \& Hernquist 1996)
complicates statistical measures of 
interaction rates.  Furthermore, 
the most dramatic, hence easily recognized, 
effects occur only in the latest stages of merging, where selection of
uniform samples can be difficult because the progenitor galaxies are
no longer resolved by standard techniques for identifying galaxies.

The first stages of a tidal encounter may induce star formation and/or
distortions in the morphologies and rotation curves of the progenitor
galaxies.  The pioneering work of Larson \& Tinsley (1978) 
explores the effects of interaction
on the colors of galaxies in pairs.
Kennicutt et al. (1987) analyze a complete set of 50
galaxies in pairs and an additional set of 32 galaxies selected
on the basis of tidal distortion.  They find an enhanced star formation
rate in interacting galaxies, relative to a field sample.  
Kennicutt et al. apply starburst models
constrained by the $\bv$ color and EW(H$\alpha$) of each galaxy
and conclude that the bursts have weak strengths and durations 
of $\sim 10^7$ years.  
Other studies of pairs of galaxies employ optical spectra, 
radio observations, infrared and optical photometry to study similar issues
(Hummel 1981;
Condon et al. 1982;
Joseph et al. 1984;
Kennicutt \& Keel 1984;
Lonsdale et al. 1984;
Keel et al. 1985;
Bushouse 1986;
Madore 1986;
Jones \& Stein 1989;
Sekiguchi \& Wolstencroft 1992;
Liu \& Kennicutt 1995a, 1995b;
Keel 1993, 1996;
Donzelli \& Pastoriza 1997;
Gao \& Solomon 1999).

Observational studies of the connections between
the star formation rate, nuclear activity, and tidal encounters 
often use samples that are small or biased towards systems with
certain kinds of morphological distortion.  
Here, we analyze new spectra of 502 galaxies in nearby pairs and n-tuples,
originally selected only on the basis of proximity in redshift space.
We use this sample to explore some of the effects of interaction
on the emission properties of galaxies in tight systems.  

The data indicate that the ages of new bursts of star formation correlate
with the spatial and velocity separations of galaxies in the same pair.
These correlations suggest that a close pass initiates a starburst,
which ages as the galaxies continue in their orbits. The Mihos \& Hernquist (1996) 
n-body/SPH models of major mergers provide a physical description of the initiation
of these bursts.
In Secs. 2 and 3, we describe the sample and our data reduction procedures.
In Sec. 4, we classify the emission-line spectra based on the source of
ionization in their nuclei and compare their properties to a well-studied
population of galaxies rich in interactions and mergers, the luminous
infrared galaxies (LIGs).  In Sec. 5, we describe the numerical simulations 
of Mihos \& Hernquist (1996; MH96 hereafter), which offer an 
interpretation of the 
differences between our pair sample and the LIGs.  This model effectively
predicts correlations between the equivalent width of H$\alpha$ (and
other lines) and the observable orbit parameters of paired galaxies; 
we describe the correlations we measure in Sec. 6.  In Sec. 7 we 
use the starburst models of Leitherer et al. (1999; L99 hereafter) 
to construct a picture where the correlation is the result of a burst of star
formation which continues and
ages as the galaxies move apart after the first pass.
This picture, applied to an even larger sample, would yield important 
constraints on the frequency and initial mass function of tidally-triggered
starbursts, along with orbital constraints on the participating pairs.
In Sec. 8 we apply consistency checks to this
interpretation of the data.  We discuss the interpretation further in 
Sec. 9 and conclude in Sec. 10.

\section{The Sample}

Based on projected spatial and velocity
separation, we extract a complete sample of 786 galaxies in
pairs and n-tuples from the original 
CfA2 redshift survey, with m$_{\rm Zwicky} \leq 15.5$.
We re-measured spectra of 502 of these galaxies; the new spectra
are more uniform than the original CfA2 data, and generally have
higher signal to noise.  
The original CfA2North covers the declination range $8.5^\circ \leq 
\delta \leq 44.5^\circ$ and right ascension range 
$8^{h} \leq \alpha \leq 17^{h}$ (B1950) and includes 6500
galaxies (Geller \& Huchra 1989; 
Huchra et al.  1990; Huchra, Geller, \& Corwin 1995). 
The original CfA2South covers the region $-2.5^\circ \leq 
\delta \leq 48^\circ$ and 
$20^{h} \leq \alpha \leq 4^{h}$ and includes 4283 galaxies
(Giovanelli \& Haynes  1985; Giovanelli et al. 1986; 
Haynes et al. 1988; Giovanelli \& Haynes 1989;  
Wegner, Haynes \& Giovanelli 1993; 
Giovanelli \& Haynes 1993; Vogeley 1993).  

The full pair sample includes 786 galaxies in 305 pairs and 51
n-tuples, originally selected with line-of-sight 
velocity separations $\Delta V \leq 1000$~km/s, projected separations 
$\Delta D \leq 50$~h$^{-1}$~kpc, and 
$cz \geq 2300$~km/s.  The 2300~km/s limit excludes 
the Virgo cluster and limits the angular sizes of the galaxies. 

Falco et al. (1999) refine the original CfA2, clearing
up problems in the catalog, which often caused deletion of
galaxies in tight systems like the ones we study.  
Although we defined our sample before Falco et al. (1999)
completed their refinement, we check our catalog against theirs
to test for missing systems.
The Falco et al. (1999) catalog contains 
146 additional systems of $N \geq 2$, including a total of
311 galaxies. The catalog also
contains 18 additional members of systems which we have
defined. Thus, we miss 329 galaxies in pairs or n-tuples, 
implying that our full sample of 786 galaxies is 70\% complete 
and that our spectroscopic sample is 45\% complete (502/1115).

With updated coordinates and velocities based on our new
spectra, 36 of the 502 galaxies with new spectra
do not fit the original selection criteria. Nonetheless, we include
these systems in the analysis.  Most come close to satisfying
the original criteria; 
all of the galaxies in the spectroscopic sample we describe
here satisfy $\Delta D \leq 77 {\rm\ h}^{-1}$~kpc and
$\Delta V \leq 1035$~km/s.

The sample consists chiefly of galaxies in pairs.
76\% of the 502 galaxies are in pairs, with two of these
in partial pairs. 14\% are in groups of $N = 3$ (22 
complete groups, one incomplete).  The remaining 
10\% are in groups of $N \geq 4$ (9 complete
groups, 3 incomplete groups).

Our sample of 502 spectra includes
the complete CfA2South pair sample from the original CfA2 
survey (310 galaxies) and 192/474
(41\%) of the CfA2North pair sample from the original CfA2North
survey. 
Because it is complete with respect to
the original CfA2 survey, the CfA2South sample
is an unbiased subset of pairs in the CfA2 survey. 

Because of our procedure for identifying candidates to include in the
spectroscopic sample, the incomplete CfA2North subsample of 192 galaxies
is biased towards galaxies with emission lines.  A K-S test of the
EW(H$\alpha$) distributions indicates a  probability of
0.013  that the CfA2North and CfA2South 
subsamples were drawn from the same distribution.  Similarly, the
$\Delta D$ and $\Delta V$ distributions differ at the
0.012 and 0.031 levels, respectively, probably because of the
bias towards line-emitting galaxies.  However, when we consider 
only the line-emitting galaxies in the CfA2North and CfA2South
samples, the differences in the $\Delta D$ and EW(H$\alpha$)
distributions disappear, with K-S probabilities of
0.20 and 0.38, respectively.  The differences in the
$\Delta V$ distributions remain, possibly because 
the $\Delta V$ distributions 
are sensitive to unavoidable large-scale structure effects.
We conclude that the CfA2North sample
is not biased with respect to the set of all line-emitting galaxies 
in pairs, although it lacks galaxies without emission lines.
We summarize these statistics in Table~\ref{tab:se}. 

\section{Observations and Data Reduction}

We use the FAST spectrograph at the 1.5m
Tillinghast reflector on Mt. Hopkins to measure spectra 
for the central regions of the galaxies.  
We use a grating with 300 lines/mm to disperse the light 
into the wavelength range $4000-7500$ \AA;  typical
exposure times are 10~--~20 minutes.  
We use standard IRAF procedures for flat-fielding and 
wavelength calibration. We measure
radial velocities from the data using the XCSAO program in IRAF 
(Kurtz \& Mink 1998),  which applies the cross-correlation 
technique of Tonry \& Davis (1979).

For this study, we extract apertures along the 3$\arcsec$-wide
slit which include the majority of the light from the galaxy.  
The aperture length ranges from 1.74$\arcsec$ to 29.7$\arcsec$, or
a projected length of 0.24~--~13.8 h$^{-1}$~kpc at the galaxy, 
with a median of 2.3 h$^{-1}$~kpc; thus, these spectra typically include
more, but not much more, than just the nuclear light.

We flux-calibrate each night using standard stars (Massey 
et al. 1988; Massey \& Gronwall 1990). 
However, not every night was cloud-free; thus we have only relative 
calibration across each spectrum.

Because we analyze full-aperture spectra, galaxy rotation and/or
outflows often lead to significantly non-Gaussian line profiles.  
Therefore, we measure line fluxes and equivalent 
widths by (1) fitting the profiles in IRAF with 
noao.onedspec.splot, (2) adding the flux within 1.5-$\sigma$ (HeI only)
to 3.5-$\sigma$, of the resulting fit, depending on the line, 
and (3) correcting for Gaussian wings (0.02 -- 13.4\% correction).
We compute the equivalent width as the ratio of the
flux in the emission line to the surrounding continuum.
We base errors in the measurements on photon statistics; a check
of the EW(H$\alpha$) errors based on repeat measurements suggests that they
are correct to within a factor of $\sim$2, except for some
outliers due to non-Gaussian effects.

Based on the measured H$\alpha$ to H$\beta$ ratio and an assumed 
intrinsic H$\alpha$/H$\beta$ = 2.85, 
we correct the line fluxes for reddening due to dust.  
If H$\alpha$/H$\beta < 2.85$, we apply no correction.
We also correct for stellar absorption;
we estimate the amount of absorption by taking the maximum 
absorption equivalent width in H$\delta$ or H$\gamma$,
then adding that equivalent width to both H$\alpha$ and H$\beta$.

\section{Spectral Classification}

Several authors uncover a high frequency of 
AGN among the most IR-luminous galaxies (Sanders et al. 1988;
Allen et al. 1991; Veilleux et al. 1995).  
Veilleux et al. (1995) analyze optical spectra of
luminous infrared galaxies (LIGs), with 
L$_{\rm ir} > 3 \times 10^{10}$L$_{\sun}$.  Almost all of their spectra, 
which are limited to the inner $\sim 1.5$~h$^{-1}$ kpc surrounding
the nucleus, have significant emission lines which allow
classification.  The AGN incidence increases with 
infrared luminosity --- the AGN fraction (Seyferts and LINERS)
of the sample for 
$\log(L_{\rm ir}/L_{\sun}) < 11$ is 29\% 
(4\% Seyferts);  for the ultraluminous infrared galaxies (ULIRGs),
with $\log(L_{\rm ir}/L_{\sun}) > 12$ the fraction 
is 62~\% (33\% Seyferts).  

Morphologies typical of late-stage interactions, including
double nuclei and tidal features, are common in 
IR-luminous galaxies.  The fraction of ULIRGs that are 
late-stage mergers may be $\gtrsim 90\%$ 
(e.g., Sanders \& Mirabel 1996;
Borne et al. 1999).  
Thus, Veilleux et al. (1995) provide a sample rich in late-stage
mergers.  The sample is a benchmark for comparison with our pair sample, 
where we expect tidal interactions associated with the early stages 
of merging.
Here, we classify our pair spectra to measure the frequency
of AGN.
With one of the empirical tests used by Veilleux et al.,
based on emission-line ratios,
we classify the objects as AGN or HII-region galaxies.  

To determine the source of ionization in the nuclear region,
we follow Baldwin, Phillips, \& Terlevich (1981) and Veilleux
\& Osterbrock (1987), who compare the line ratios 
[NII]($\lambda$6584)/H$\alpha$ 
and [OIII]($\lambda$5007)/H$\beta$.
This test, along with similar tests based on [OI] and [SII],
works because x-ray photons from an AGN produce an extended, 
partially ionized region around the source where substantial
collisional excitation of N$^{+}$ (and others) occurs;
objects with larger [NII]($\lambda$6584)/H$\alpha$ ratios are
probably AGNs (Veilleux \& Osterbrock 1987).  O$^{++}$
ions are produced by lower-energy UV photons; 
[OIII]($\lambda5007$)/H$\beta$ is more of a 
measure of the UV flux close to the source 
(Veilleux \& Osterbrock 1987). Veilleux \& Osterbrock (1987) calibrate 
the test empirically, using
known Seyferts, LINERS and HII-region galaxies.

\begin{figure}
\plotone{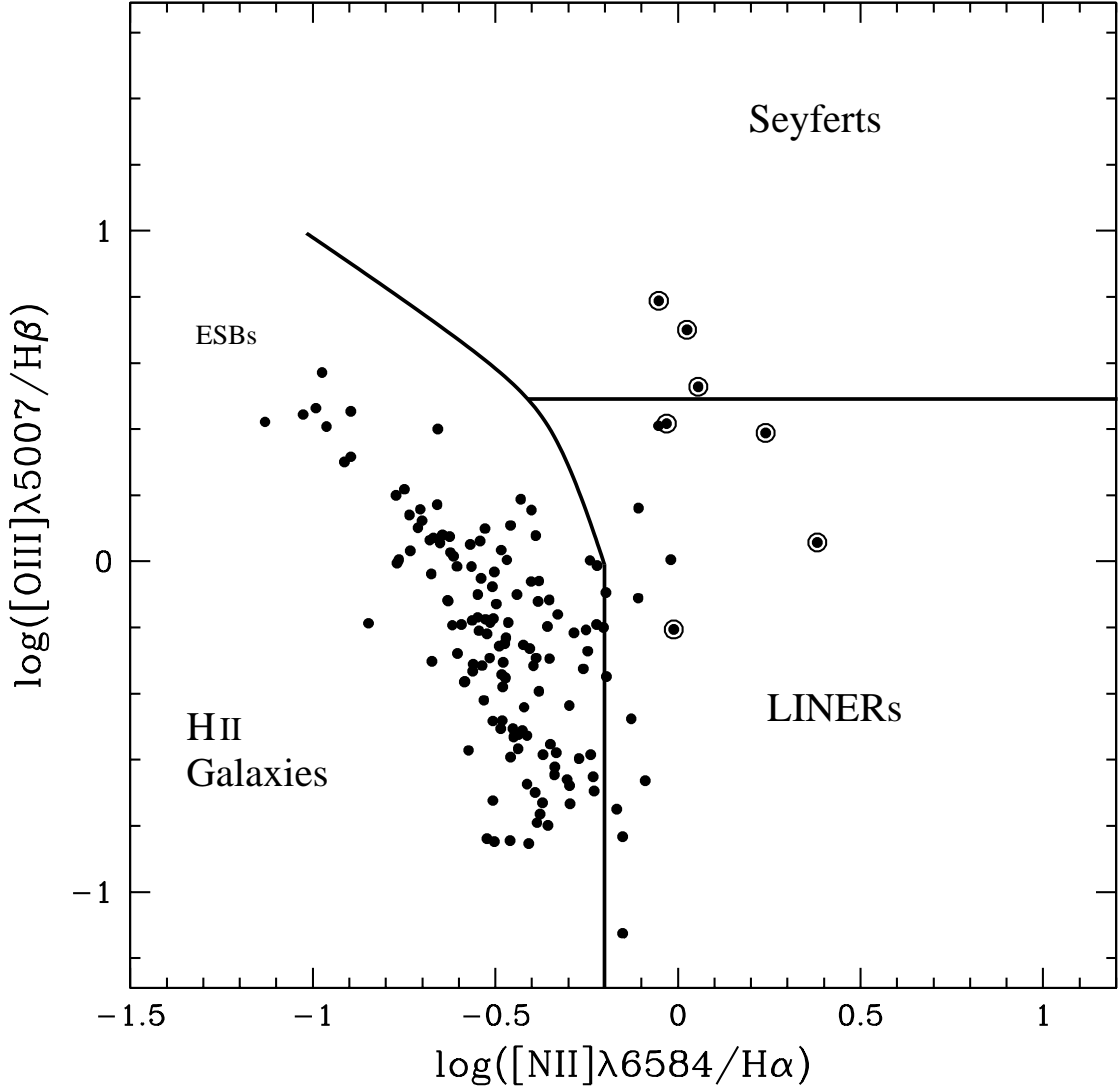}
\caption{log([NII]$\lambda6584$/H$\alpha$) vs. [OIII]$\lambda5007$/H$\beta$
for emitting galaxies in our sample.  Solid lines divide the 
sample into different types based on the empirical results 
of Veilleux \& Osterbrock (1987).
Circled points are galaxies with broad emission lines typical of
Seyferts; their ratios are uncertain due to deblending problems.
We label the population of $\sim$8 extreme starburst galaxies as ESBs.}
\label{fig:vo1}
\end{figure}

Fig.~\ref{fig:vo1} shows flux ratios for the 
diagnostic lines in the 149 sample galaxies
with significant H$\alpha$, H$\beta$, [OIII] and [NII] emission.
The solid lines divide the sample into
Seyferts, LINERs or HII-region galaxies according to
Veilleux \& Osterbrock (1987).  We correct the ratios for
Balmer absorption and reddening (Sec.~3).  
Twenty galaxies have broad emission lines chiefly due
to nuclear activity, although in a few cases, the broad lines 
may be the result of rapid rotation and extended 
emission.  Galaxies with broad lines
have very uncertain line ratios due to blending; we plot the 7 
with significant [OIII] as larger, partially-filled 
circles.  Some of these galaxies are probably
Seyferts even though they do not 
fall in the Seyfert region in Fig.~\ref{fig:vo1}.

The incidence of AGN is much smaller in our sample
than in the LIG sample of Veilleux et al. (1995).
Only 3 of the classifiable galaxies clearly 
have line ratios typical of Seyferts. One additional 
galaxy, NGC 2622 (not plotted in 
Fig.~\ref{fig:vo1} due to
heavily blended lines)  is a known Seyfert 1.8,
bringing the total to 4/150 (2.7\%) galaxies with Seyfert-type 
line ratios.  An additional 15/150 (10.0\%) galaxies fall 
in the LINER range on the plot; many of these galaxies
are near the HII-galaxy border.  The total AGN fraction 
of the classifiable sample is 19/150 (13\%), compared
with 41\% of LIGs (Veilleux et al. 1995).

In fact, the incidence of Seyferts in pairs 
is close to the incidence of Seyferts in the field.
Of the complete CfA2South region, 
only 2/310 (0.6\%) have detectable 
Seyfert-type line ratios -- this percentage is similar to
the fraction of Seyferts in the bright portion of
the CfA redshift survey, 1.3\% (Huchra \& Burg 1992). 
In our sample, 8/310 (2.6\%) have LINER-type line ratios
 with observable [OIII]($\lambda$5007)/H$\beta < 3$.  
Ho, Filippenko \& Sargent (1997) report a LINER frequency 
of 19\% in the field, although their sample is not ideal
for comparison because they detect emission at very 
low levels in their high-quality spectra of a large sample of 
nearby galaxies rich in low-luminosity dwarfs.  In any case,
we detect no clear excess of LINERs or AGN in our emission-line
sample.

The differences between our results and the results
of Veilleux et al. could arise because of 
differences in reduction procedures.
Veilleux et al. (1995) note that the star-forming region surrounds 
the active nucleus in their sample galaxies.
This emission might conceivably wash out the AGN signature in our
full-aperture spectra, typically $\sim$2 times as wide as
the apertures they extract. 
However, when Veilleux et al. (1995)
double the aperture size for
23 galaxies with major-axis spectra none of their classifications
change from AGN to HII-galaxy.  Errors in the correction
for Balmer absorption move points vertically,  more than
horizontally, in Fig.~\ref{fig:vo1}.  
Thus, circumnuclear starbursts (see e.g., Delgado \& Heckman 1999)
appear to affect the two samples
in the same way, and differences between our methods of
correction for Balmer absorption are probably unimportant.
Thus, the excess of AGN in the Veilleux 
et al. infrared luminous sample appears to be truly
absent from our sample and is not an artifact of differences
in reduction technique.

The sample of Veilleux et al. (1995) does not contain
the ``extreme starburst galaxies'' 
(Allen et al. 1991) which populate the upper left corner of
Fig.~\ref{fig:vo1} in our sample.  
Veilleux \& Osterbrock note the absence of these 
high-ionization HII galaxies in their sample and the apparent dearth
of these objects in other LIG samples (Leech et al. 1989;
Armus, Heckman, \& Miley 1989; Allen et al. 1991; 
Ashby, Houck \& Hacking 1992).
Allen et al. (1991) report a small number of these
galaxies at IR luminosities, both lower and higher
than the Veilleux et al. cutoff. 
These galaxies are less numerous than Seyferts
in their sample, whereas they outnumber the Seyferts significantly
in ours.

Both the lack of AGNs and the presence of a
population of the ``extreme starburst galaxies''
suggest that we are sampling a population of objects which
differs from the luminous infrared galaxies, and
from the IRAS sources in general.  The difference could
be a result of both the pair selection and 
the optical (blue) selection of our sample:
IR selection may favor late-stage mergers, dusty mergers, and merger
remnants, whereas blue selection may favor younger star-forming
systems.  Furthermore, the formation of an active nucleus
in the LIGs appears to be linked to the latest stages of the merging
process --- Veilleux et al. (1995)
report that on average, the AGN systems are morphologically 
more advanced mergers than the HII-region galaxies.

On the basis of the observed lack of AGN in our sample (relative to the
LIGs), and numerical simulations which track gas
infall in interactions (Mihos \& Hernquist 1996),
the evidence points toward a burst of star formation which
{\it precedes} the final merging stages, including the 
ULIRG phase, and/or the AGN phase, in some 
merging galaxies.  We pursue the idea that the initial tidal 
interaction triggers the burst.
Next, we review the elements of this picture.

\section{Simulations of Starbursts in Pairs}

N-body/SPH simulations (e.g., Hernquist \& Mihos 1995; 
Mihos \&  Hernquist 1996)  show that interactions between 
disk galaxies without bulges (i.e. with shallow central potentials)
can trigger a burst of star formation before the 
final merger, when the progenitors still appear as a resolved 
galaxy pair.

Mihos \& Hernquist (1996; MH96) investigate star formation in 
major mergers with hybrid n-body/SPH simulations. 
They model star formation in the gas according to a Schmidt (1959) 
law, with the star formation rates (SFR)~$\ \propto\ \rho^{\rm -n}$, where
$\rho$ is the density of gas; they also include a prescription for
feedback into the interstellar medium.  They incorporate
hybrid particles which turn from gas into stars during the 
simulations. MH96 find that interactions trigger 
star formation after the first pass and again during the merger.
The first orbit is fairly elongated, with 
r$_{\rm peri} \sim 9 $~kpc and r$_{\rm apo} \sim 42$~kpc, 
scaled to the Milky Way, where r$_{\rm peri}$ and r$_{\rm apo}$ are the
galaxy separations at perigalacticon and apogalacticon,
respectively. The galaxies merge after a second, much smaller,
orbit with r$_{\rm peri} \sim 5$~kpc and r$_{\rm apo} \sim 9$~kpc (MH96).

In MH96 models with bulges, the bulge suppresses the early 
(post first pass) star formation because the
steep central potential associated with the bulge inhibits 
the formation of a central, non-axisymmetric mass perturbation, a
bar.  When the final merger occurs, most of the gas has not
been turned into stars --- a burst of star formation occurs 
during the merger, which MH96 identify with the ULIRG phase.
However, in the MH96 models without 
bulges, a bar forms as a result of a close pass, long before
the final merger.  Gravitational
torques from the bar allow gas to stream toward the nucleus 
and initiate a starburst which begins $\sim$100 Myr after the first 
close pass.  
In these systems, the final merger produces a
weaker starburst, and presumably no ULIRG,
because the earlier burst depleted the available gas.  
In Sec.~7, we ask whether the starbursts observed in our 
sample could have formed in an initial pass, before the final
merger.

The MH96 simulations cover a small region of parameter
space, including only interactions with small ($\sim$9~kpc)
impact parameters on the first close pass.  However, Barton,
Bromley, \& Geller (1999a), show that even much wider passes,
with impact parameters greater than $\sim$50~kpc, 
perturb the galaxies kinematically, 
forming barlike structures which could allow gas infall to
initiate a (possibly weaker) starburst.  More simulations are necessary 
to measure the magnitude of the induced starburst as a 
function of impact parameter, halo potential, inclination
angle, etc.  However, the simulations to date
suggest: (1) star formation may be triggered during a 
close pass, due to torques from distortions in the host
galaxy; this starburst is visible while the galaxies are
a resolvable pair, (2) a starburst which occurs as a 
result of a close pass will be delayed until some time
after the close pass, while the gas falls in and collects in the center
of the galaxy.  This burst then continues for 
an extended period of time (a few hundred Myr), while the tidal tails
are falling back onto the galaxies. If the burst is strong,
it will deplete the gas and weaken the starburst that occurs
when the galaxies finally merge.  Finally, (3) the potential 
of the galaxy strongly affects its response to a close pass --- 
only galaxies with shallow enough central potentials 
can form bars, which 
allow gas infall.  The simulations have implications for
star formation in resolvable pairs, which we explore below.

\section{Correlations Between Star-Formation Signatures and 
Observable Orbit Parameters}

If, as suggested by MH96, 
interactions trigger the starbursts observed in 
close pairs, their spectral 
properties should depend on local conditions, 
including orbit parameters.  
Some authors measure a correlation between $\Delta D$, the
projected separation on the sky, or $\Delta V$, the
line-of-sight velocity separation, and galaxy emission-line
strengths or colors (Madore 1986;  %
Kennicutt et al. 1987, 111 galaxies;
Jones \& Stein 1989), 
or molecular gas depletion 
(Gao \& Solomon 1999, 50 mergers) in samples of close pairs
of galaxies. In contrast,
Donzelli \& Pastoriza (1997)
find no trend for 83 galaxies in pairs.

Here, we examine the correlation between the emission properties of the
galaxies in our sample
and $\Delta D$ and $\Delta V$. 
Our sample is large enough that we can begin 
to understand the chief causes 
of the correlation, if present; 
we can also separate the effects of a recent interaction 
from effects due to the large-scale environment of the pair.  

\begin{figure}
\plottwo{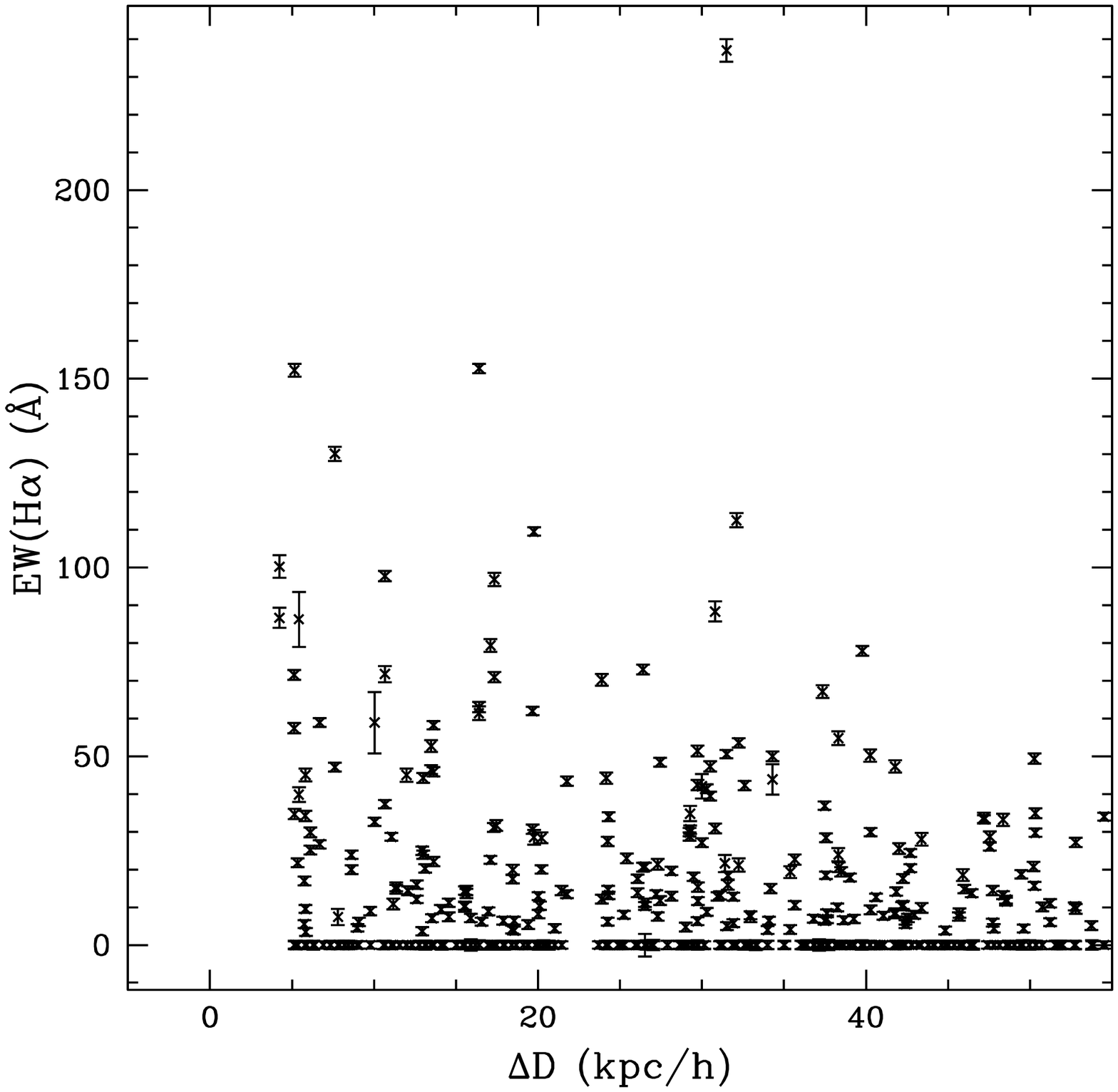}{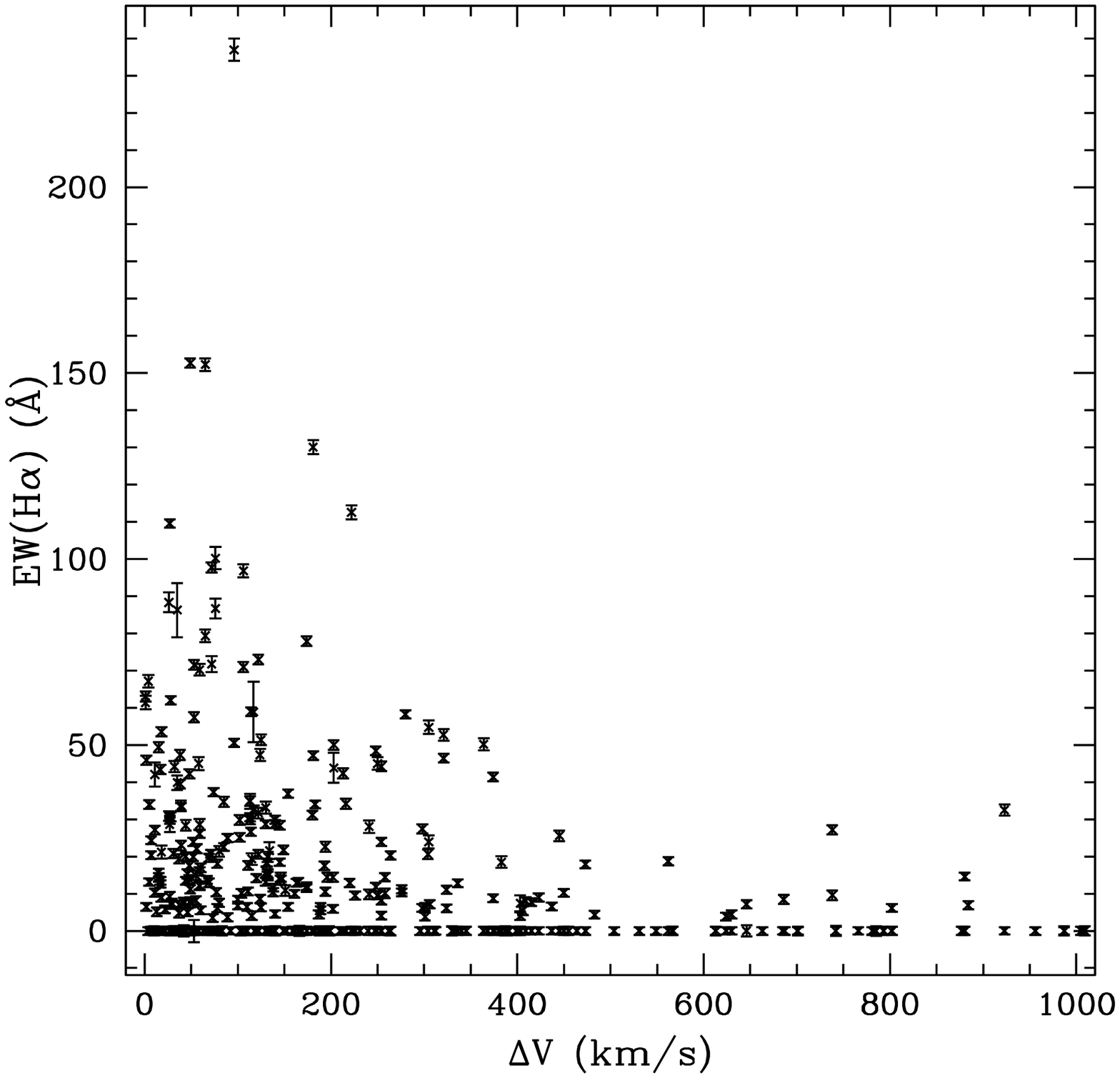}
\caption{EW(H$\alpha$) as a function of (a) $\Delta D$ (excluding
points with $\Delta D > 55$h$^{-1}$~kpc), and (b) $\Delta V$ 
(excluding points with $\Delta V > 1000$~km/s).}
\label{fig:ew1}
\end{figure}

Fig.~\ref{fig:ew1} shows EW(H$\alpha$)
for galaxies in our sample as a function of $\Delta D$ and 
$\Delta V$. We correct EW(H$\alpha$) for Balmer absorption.  We include
groups of $N \geq 3$; in these cases we set $\Delta D$ equal
to the separation from the nearest neighbor in
the n-tuple.  A trend is evident in Fig.~\ref{fig:ew1}a --- a
Spearman rank test yields only a $5.1 \times 10^{-3}$ 
probability, P$_{\rm SR}$, associated with 
the null hypothesis that the distributions are 
independent.  We apply the Spearman rank test in a robust manner,
by adding random Gaussian deviates (distributed as the errors) 
to the values of EW(H$\alpha$) and re-computing the Spearman rank
probability; we quote the median for 200 of these tests. 
The plot shows a wide
distribution of EW(H$\alpha$) for pairs with small $\Delta D$;
at larger separations, the sample contains only galaxies with 
moderate and small EW(H$\alpha$).  
Fig.~\ref{fig:ew1}b shows a similar, even stronger 
trend with $\Delta V$, with a small
P$_{\rm SR} = 6.3\times 10^{-5}$.  If we restrict the sample 
to only the complete CfA2South region, the smaller sample leads to a 
larger P$_{\rm SR}$: for
$\Delta D$, P$_{\rm SR} = 5.4 \times 10^{-2}$, and for
$\Delta V$, P$_{\rm SR} = 2.2 \times 10^{-2}$.

The galaxies span a factor of $\sim$6 in distance from us.  
In addition, some pairs in our sample are embedded within dense cluster 
environments, where gas stripping may play a role.
Therefore, we test the possibility that the correlations are a result
of large-scale effects, rather than local physics associated with the
interaction.  

\begin{figure}
\plottwo{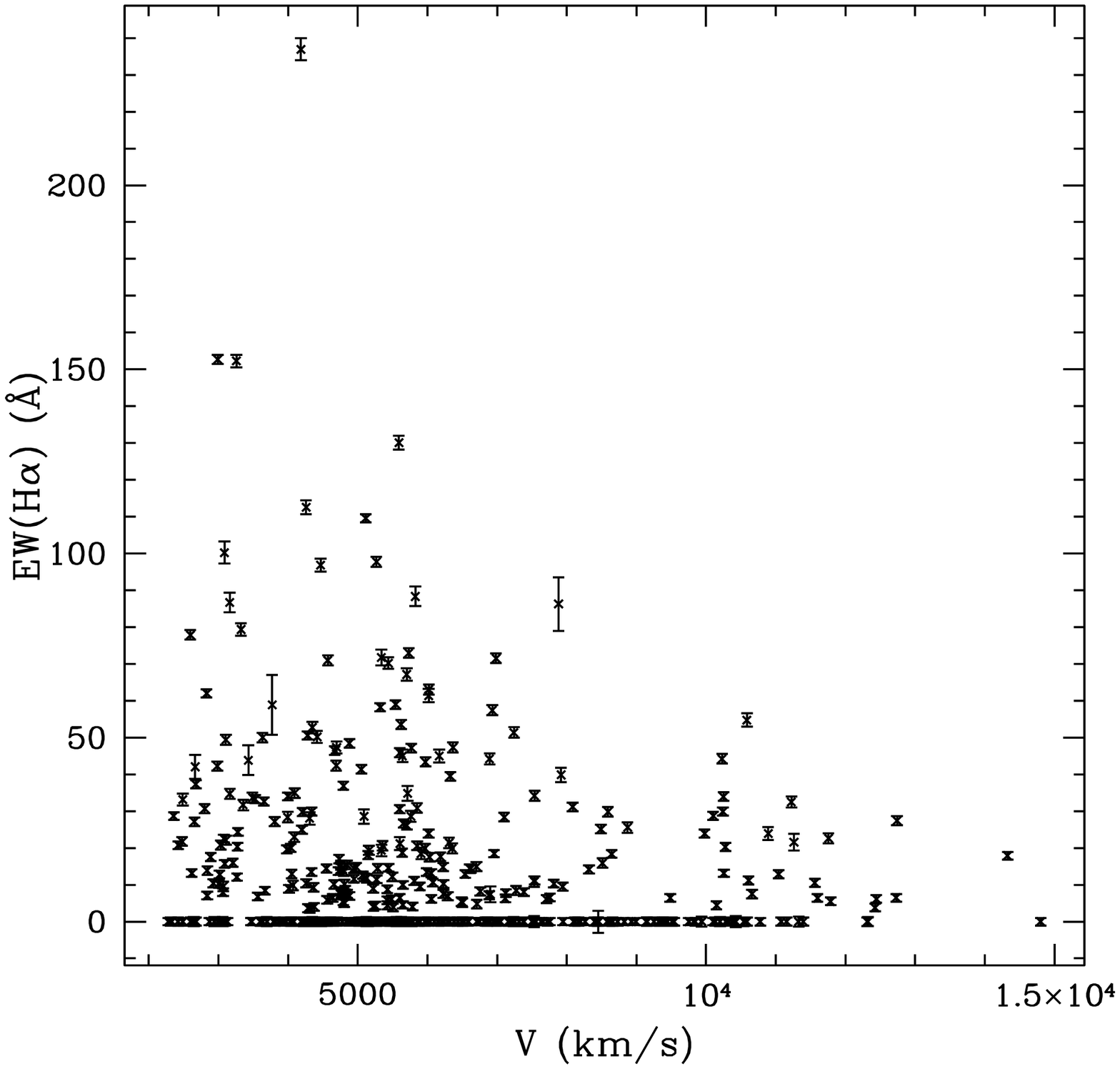}{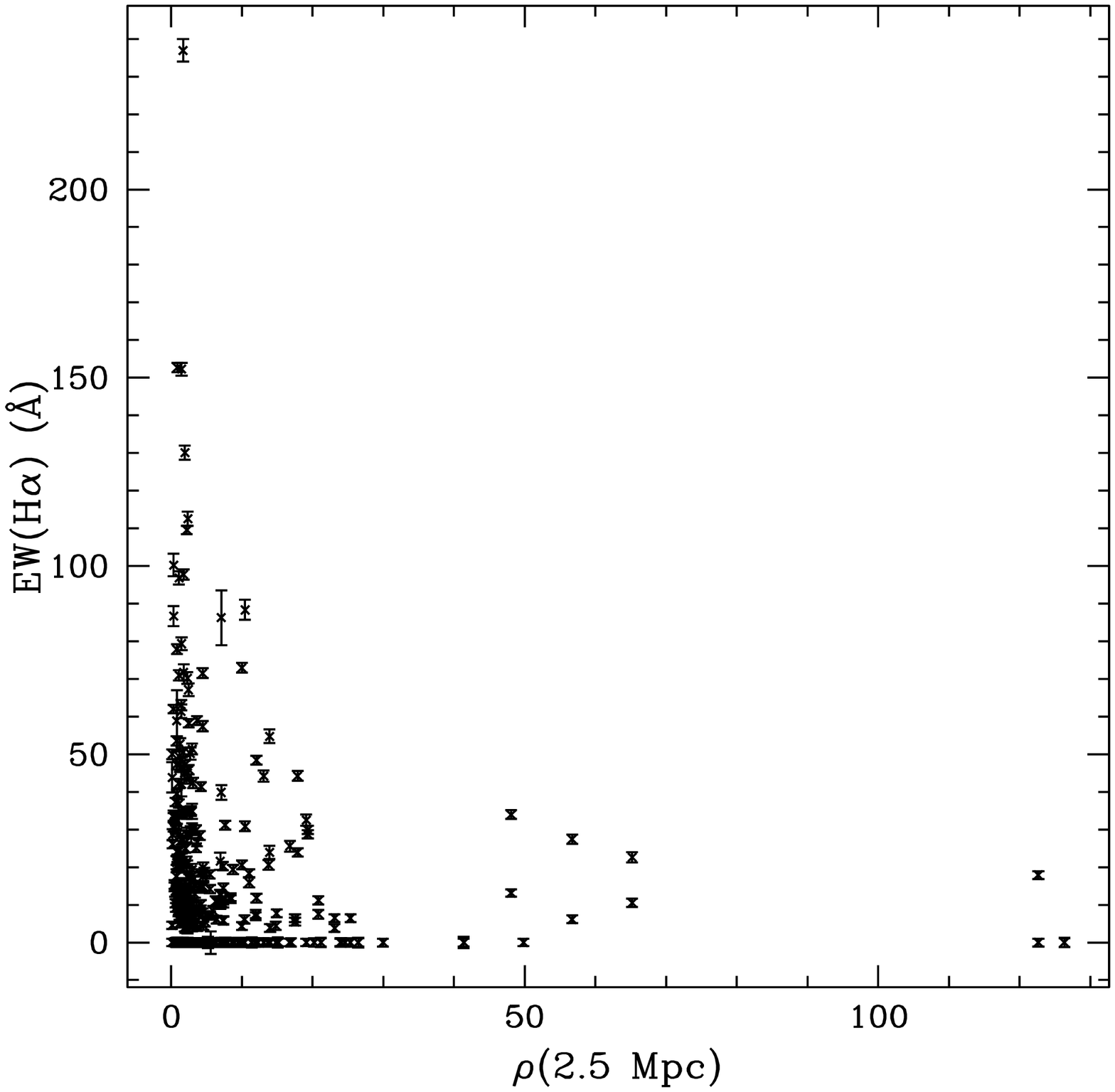}
\caption{EW(H$\alpha$) as a function of (a) systemic recession velocity, and
(b) density of the surrounding environment, smoothed to 2.5h$^{-1}$ kpc and
normalized to the survey mean.}
\label{fig:ew2}
\end{figure}

The plots in Fig.~\ref{fig:ew2} show that EW(H$\alpha$) correlates
with global parameters, including
redshift and the surrounding galaxy density smoothed on
a 2.5 h$^{-1}$ Mpc scale.  
Fig.~\ref{fig:ew2}a shows EW(H$\alpha$) as a function of
redshift (P$_{\rm SR} = 6.3 \times 10^{-3}$). 
The correlation may arise because nearby we sample 
intrinsically faint galaxies which tend to have large EW(H$\alpha$)
relatively more frequently.

For Fig.~\ref{fig:ew2}b, we measure the surrounding density using the 
smoothed-density estimator of Grogin \& Geller (1998) applied to 
the CfA2 redshift survey, with a smoothing scale of 2.5~h$^{-1}$Mpc.  
They normalize this smoothed density, $\rho_{2.5}$, to a mean survey density 
of 1.  The correlation between EW(H$\alpha$) 
and $\rho_{2.5}$ is extremely strong; the Spearman rank 
probability of no correlation is only $1.5 \times 10^{-9}$.
The strength of this correlation with density
(see e.g., Hashimoto et al. 1998 for a recent analysis), 
reflects, in part, the well-known morphology-density 
relation (Dressler 1980).
The correlation 
has implications for all comparisons of 
emission properties of pair and compact 
group galaxies to those of isolated galaxies.  
Because apparently tight systems are
preferentially found in dense environments,
{\it the density of the surrounding environment
must be taken into account when comparing compact systems to 
the field.}  Only systems in similar environments should
be compared directly.

The local orbit parameters ($\Delta D$, $\Delta V$) also
correlate with the global parameters ($z$, $\rho_{2.5}$).
Both $\Delta V$ and $\Delta D$ correlate with $\rho_{2.5}$;
on average, $\Delta V$ and $\Delta D$ are larger in high-density
environments.  In addition, $\Delta V$ correlates with 
redshift, in the same sense.  The systems in the 
densest environments occur exclusively at
$z \geq 8000$~km/s because of the Great Wall and other large-scale 
structures, along with sparse sampling at high
redshift.  Thus, these global/local correlations are probably 
all primarily distance-dependent and large-scale structure effects which 
result from sparse sampling at high redshift and from the
more frequent occurrence of rich clusters which have a larger
$\Delta V$ for close pairs.

We seek to isolate the effects of local interaction on the 
star-forming properties of galaxies in pairs.  Therefore, for
the sake of caution, we test the possibility that the 
$\Delta D$~--~EW(H$\alpha$) and 
$\Delta V$~--~EW(H$\alpha$) correlations are indirect, 
reflecting only the effects of sparse sampling, 
the morphology-density relation, gas stripping or other
large-scale structure effects.
In other words, we test the possibility that, for example, the
correlation results because a 
smaller $\Delta D$ implies a lower $\rho_{2.5}$ (from the
global/local parameter correlation), which implies a 
larger EW(H$\alpha$) (from the morphology-density relation)
indirectly,
as opposed to a direct $\Delta D$~--~EW(H$\alpha$) correlation
resulting from the effects of local interactions.  

\begin{figure}
\centerline{\epsfysize=7in%
\epsffile{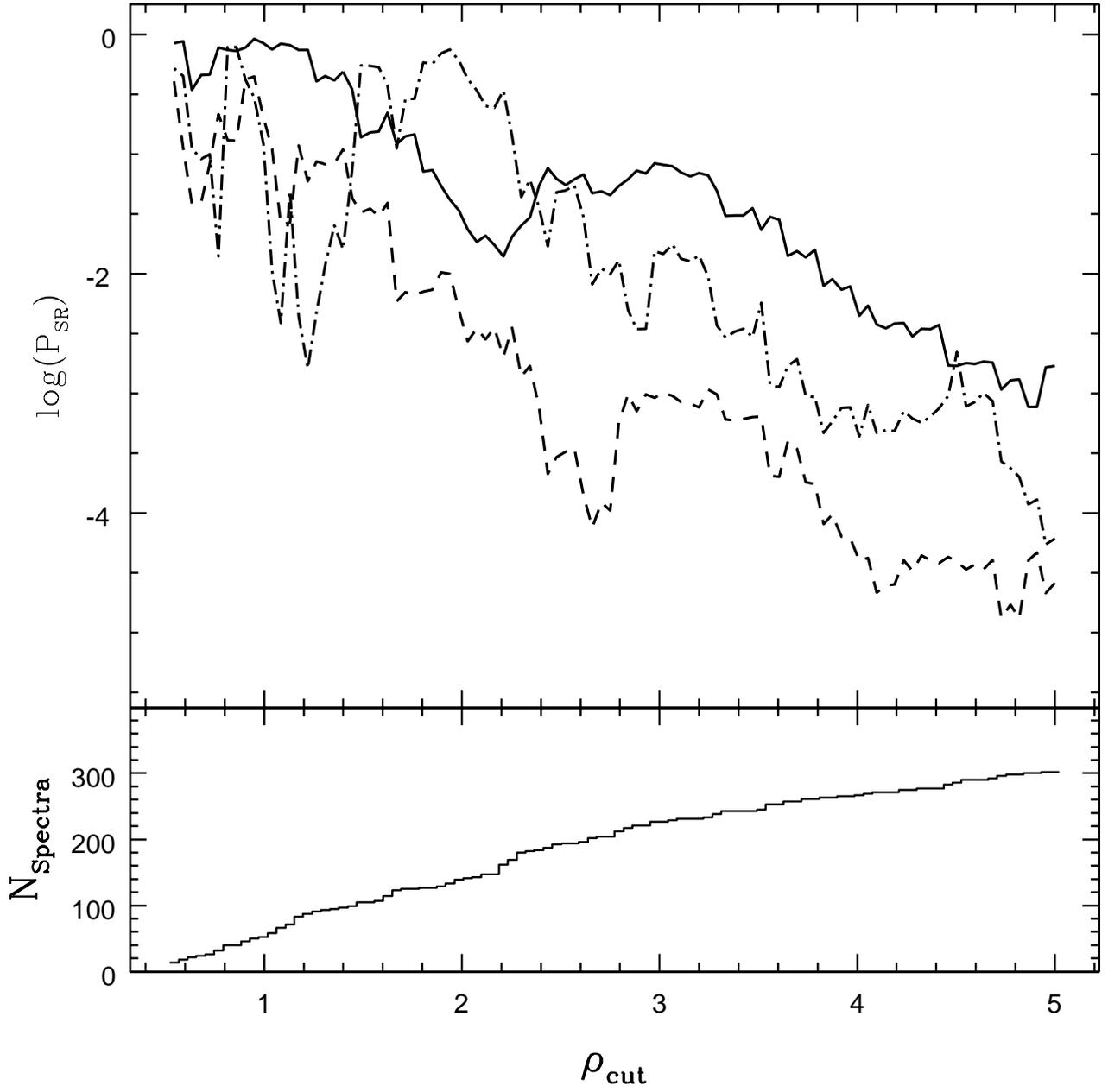}}
\caption{Spearman rank probability of no correlation for galaxies
in environments with $\rho_{2.5} < \rho_{\rm cut}$
as a function of $\rho_{\rm cut}$ for $\Delta D$ vs. EW(H$\alpha$)
(solid line), $\Delta V$ vs. EW(H$\alpha$) (dashed line) and
$\rho_{2.5}$ vs. EW(H$\alpha$) (dot-dashed line).  The bottom
panel shows the number of galaxies (with $\rho_{2.5} < \rho_{\rm cut}$)
that contribute to each point.}
\label{fig:ew3}
\end{figure}

In Fig.~\ref{fig:ew3}, we test for correlations between
global and local parameters and star formation as a function
of density cutoff, $\rho_{\rm cut}$, for the sample.  
For each point, we 
include all pairs or n-tuples with $\rho_{2.5} < \rho_{\rm cut}$, and
compute the probability of no $\Delta D$~--~EW(H$\alpha$) correlation 
(solid line), no $\Delta V$~--~EW(H$\alpha$) correlation (dashed line), 
and no $\rho_{2.5}$~--~EW(H$\alpha$) correlation (dot-dashed line).  
The lower section of the figure shows
the number of galaxies contributing to each point. 
Decreasing the number of points in a sample necessarily 
lowers the apparent strength of a correlation because it decreases its
statistical significance.  However, for densities 
$\rho_{\rm 2.5} \lesssim 2.2$, the
$\rho_{2.5}$~--~EW(H$\alpha$) correlation fades; the  
$\Delta V$~--~EW(H$\alpha$) and 
$\Delta D$~--~EW(H$\alpha$) correlations remain.  
In other words, the local dynamics appear to have a comparable or
greater effect on the galaxy emission-line properties than the global, 
smoothed $\rho_{2.5}$.  
This result is consistent with Postman \& Geller (1984), 
who find that the morphology-density relation fades for
(local) densities below 5 galaxies Mpc$^{-3}$.

\begin{figure}
\plottwo{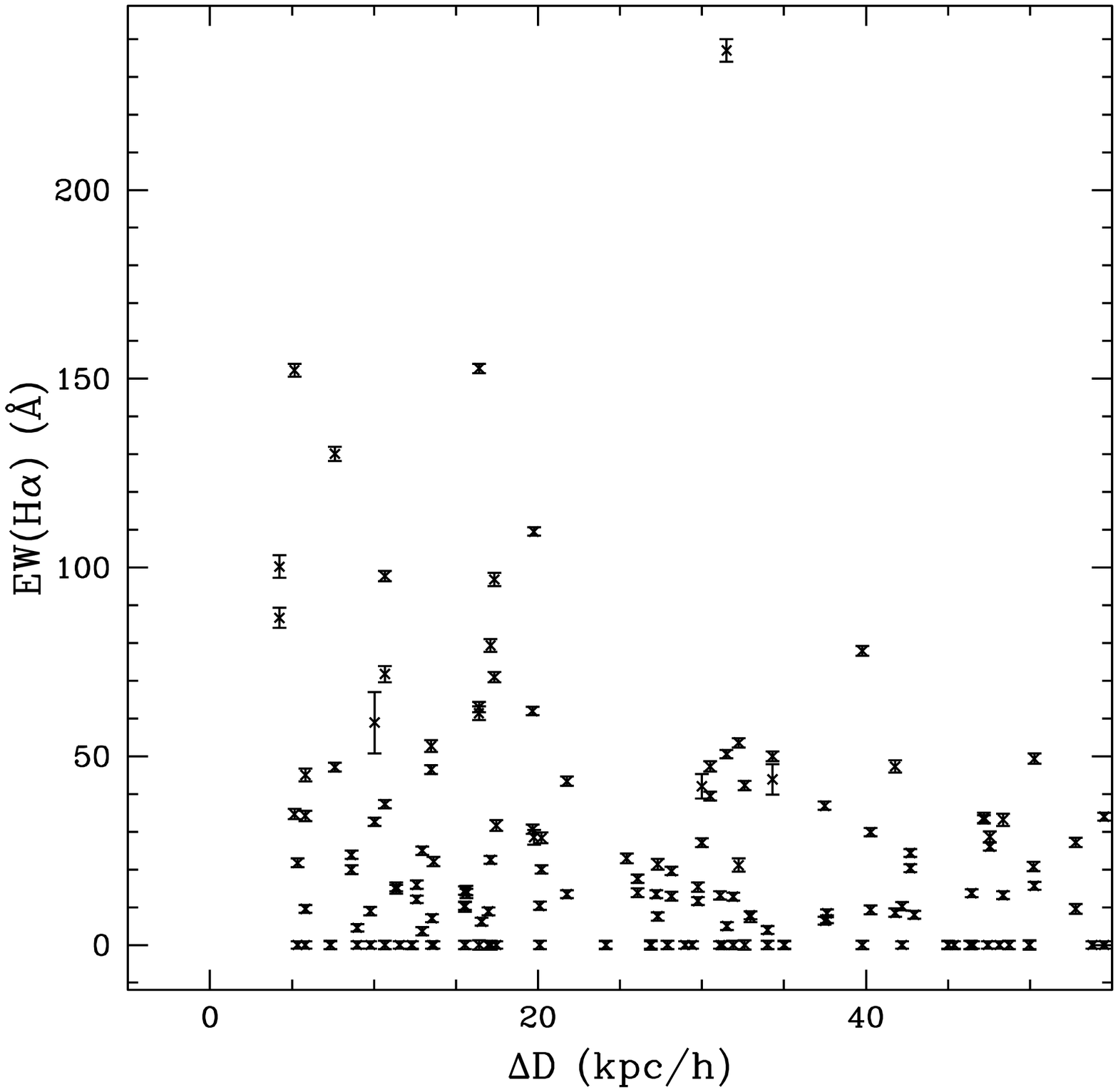}{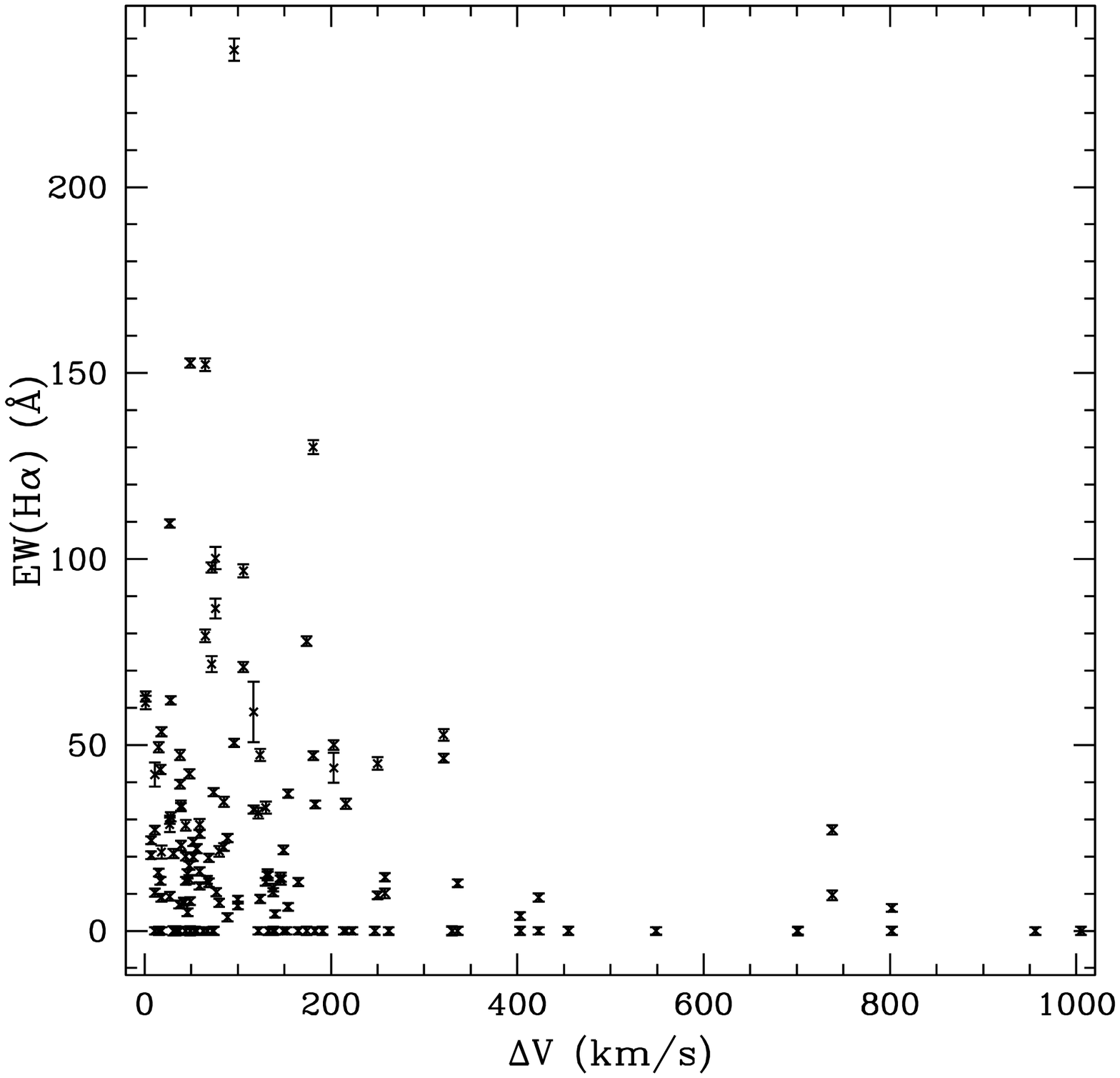}
\caption{Same as Fig.~2, restricted to galaxies in environments
with $\rho_{2.5} \leq 2.2$.}
\label{fig:ew4}
\end{figure}

Fig.~\ref{fig:ew4} shows EW(H$\alpha$) as a function
of  $\Delta D$ and $\Delta V$  for the 162 galaxies in regions
with $\rho_{2.5} \leq 2.2$.  Thus, Fig.~\ref{fig:ew4} compares only pairs in
similar, sparsely populated environments.
The $\Delta D$~--~EW(H$\alpha$) and $\Delta V$~--~EW(H$\alpha$)
correlations are apparent, with P$_{\rm SR} = 1.7 \times 10^{-2}$ 
and P$_{\rm SR} = 1.9 \times 10^{-3}$, respectively.  

The density restriction limits
the sample in redshift, reducing distance-dependent effects: 
92\% of the galaxies lie in the 
range $ 2650 $~km/s~$ \leq z \leq 6500 $~km/s.  
Thus correlations between 
local orbit parameters ($\Delta D$ and $\Delta V$) and
global parameters (redshift and $\rho_{2.5}$) fade
(except for $\Delta V$~--~redshift).  In any case, 
they cannot give rise to the $\Delta V$~--~EW(H$\alpha$)
correlation, because the 
$\rho_{2.5}$ and redshift no longer correlate with EW(H$\alpha$),
with Spearman rank probabilities of P$_{\rm SR} = 0.33$ and 
P$_{\rm SR} = 0.78$, respectively.
Therefore, we conclude that the $\Delta D$~--~EW(H$\alpha$)
and $\Delta V$~--~EW(H$\alpha$) correlations result from 
local interactions.

The strengths of any correlations between emission-line properties
and orbital parameters are necessarily reduced by the presence 
of interlopers.  We compute the number of interlopers
under the assumption of a uniform galaxy
distribution (i.e. irrespective of clustering properties), with 
$\langle \frac{\delta \rho}{\rho} \rangle 
= \langle \rho_{2.5} \rangle $, where 
$\langle \frac{\delta \rho}{\rho} \rangle$ is the overdensity 
relative to the survey mean and  $\langle \rho_{2.5} \rangle = 1.2$
is the mean value of $\rho_{2.5}$ for the pairs with $\rho_{2.5} \leq 2.2$.
We sum the expectation values
of the number of galaxies 
within 50~h$^{-1}$~kpc and 1000~km/s by integrating
the luminosity function (Marzke et al. 1994), for each of the 
original CfA2 survey galaxies between 2300 km/s and 15,000 km/s.
The expectation value of the total number of false pairs
roughly equals 26.3 pairs, or 52/786 galaxies (6.7\%).
Thus, by restricting to the 162 galaxies with $\rho_{2.5} \leq 2.2$, we
probably limit the frequency of interlopers to $\lesssim$10\%.

\begin{figure}
\plottwo{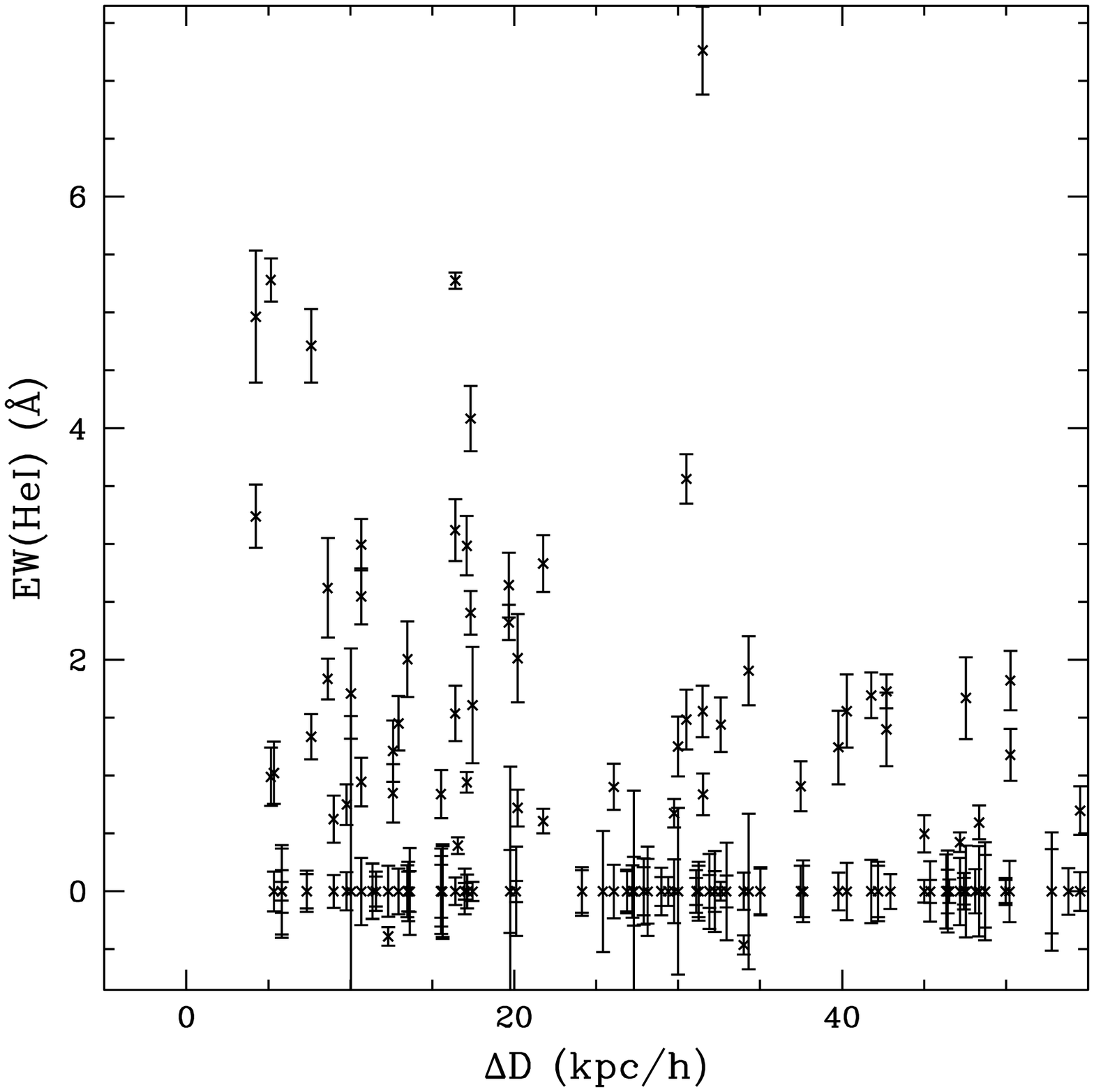}{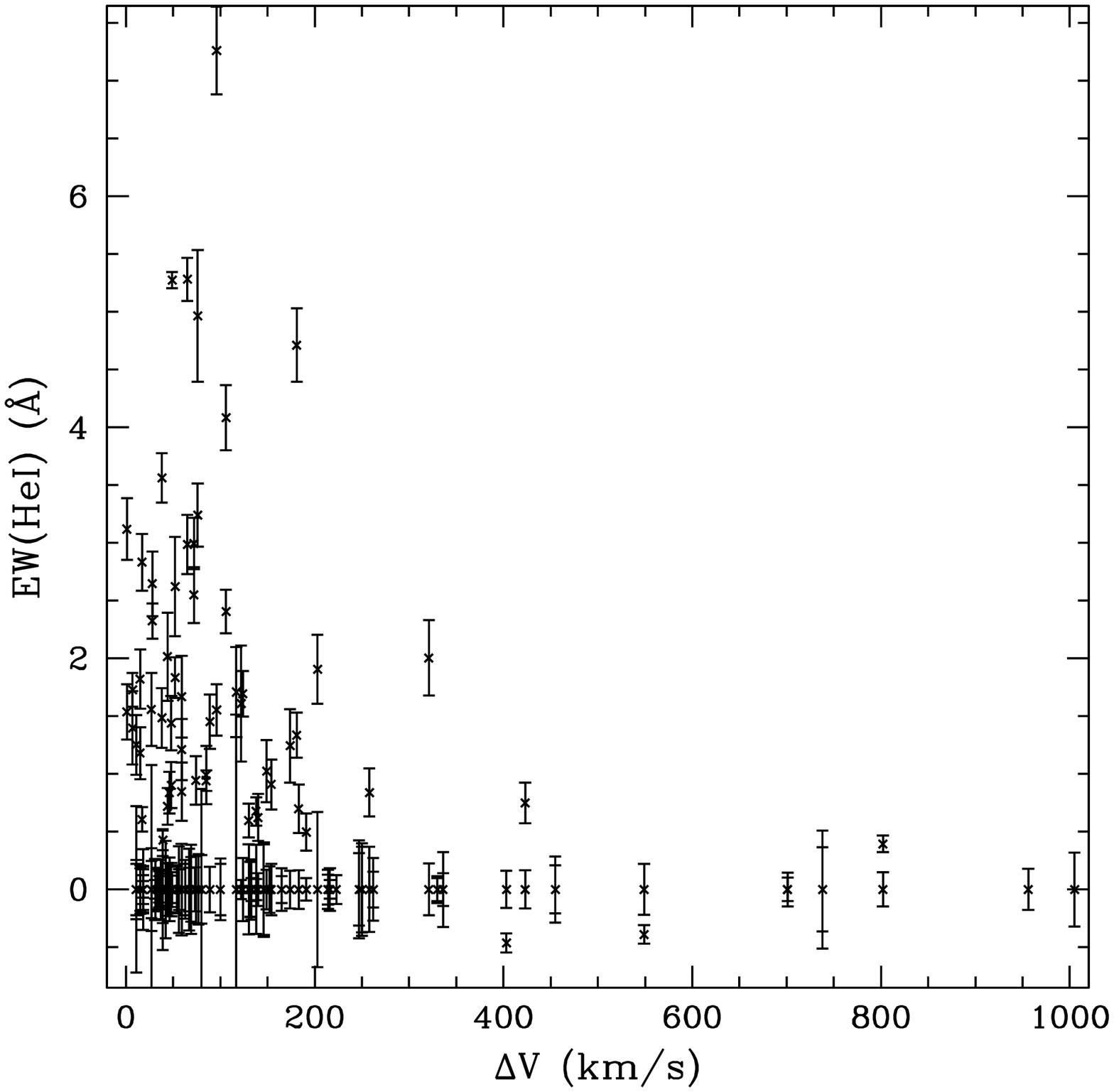}
\caption{Same as Fig.~5, except
that the ordinate is EW(HeI$\lambda5876$).}
\label{fig:ew5}
\end{figure}

HeI is a better indicator of recent star formation than 
H$\alpha$ because 
the Helium emission fades rapidly after a starburst, as the
stars hot enough to ionize Helium die (e.g., Leitherer et al.
1999; see also Sec.~8.3).  In Fig.~\ref{fig:ew5} we
plot EW(HeI) at 5876 \AA\  as a function of $\Delta D$ and
$\Delta V$ (for the restricted $\rho_{2.5} \leq 2.2$ sample). 
Fewer galaxies have detectable HeI --- 
it is a weaker line than H$\alpha$.  Spearman rank tests indicate
that the correlations are significant, with 
P$_{\rm SR} = 1.5 \times 10^{-2}$ and 
P$_{\rm SR} = 1.4 \times 10^{-2}$, for EW(HeI) vs. $\Delta D$ and $\Delta V$,
respectively.
Thus, the correlations between star formation indicators and
the observable orbit parameters are strong, even when we 
restrict the sample to the lowest-density environments. 

\begin{figure}
\plottwo{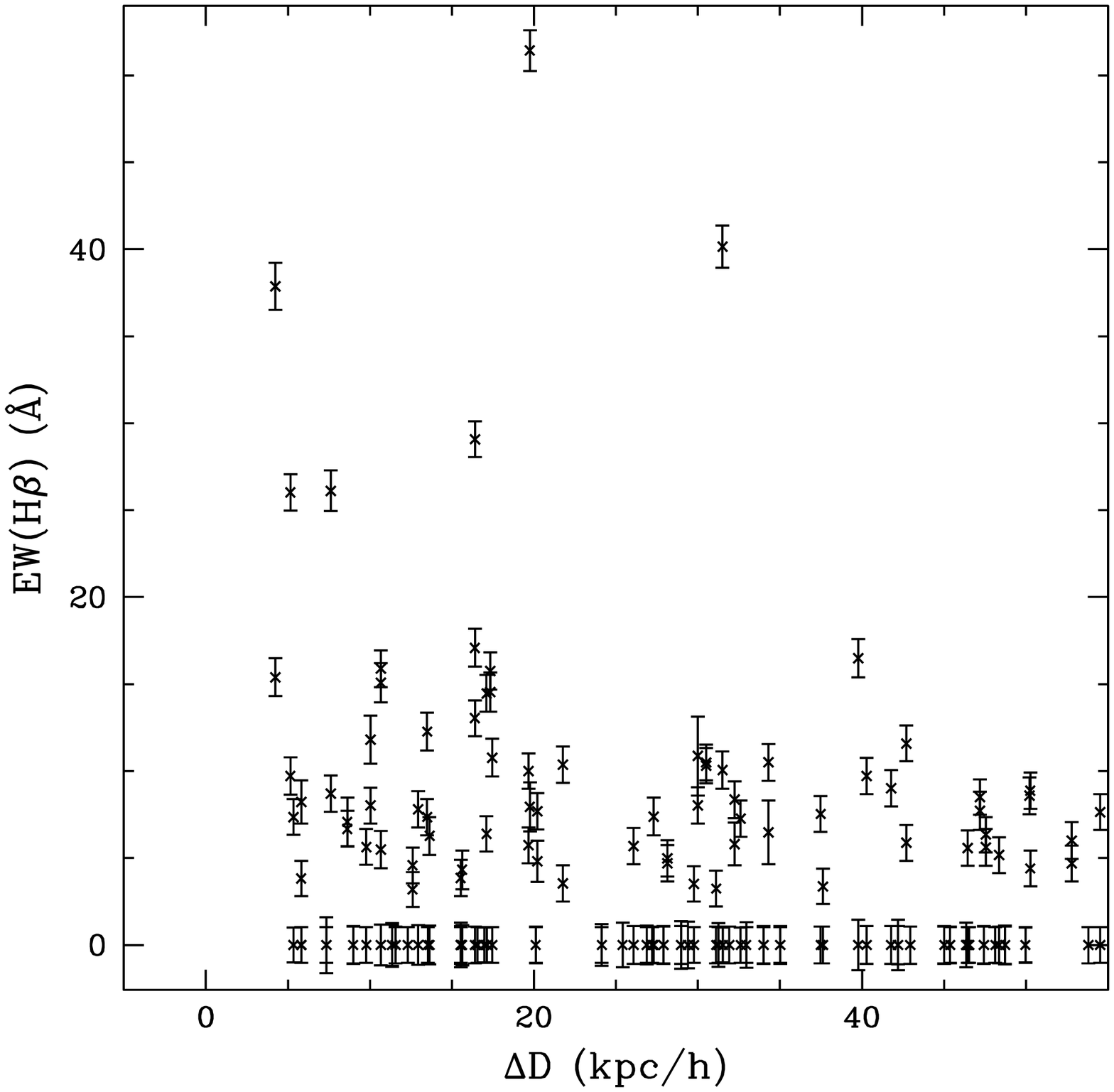}{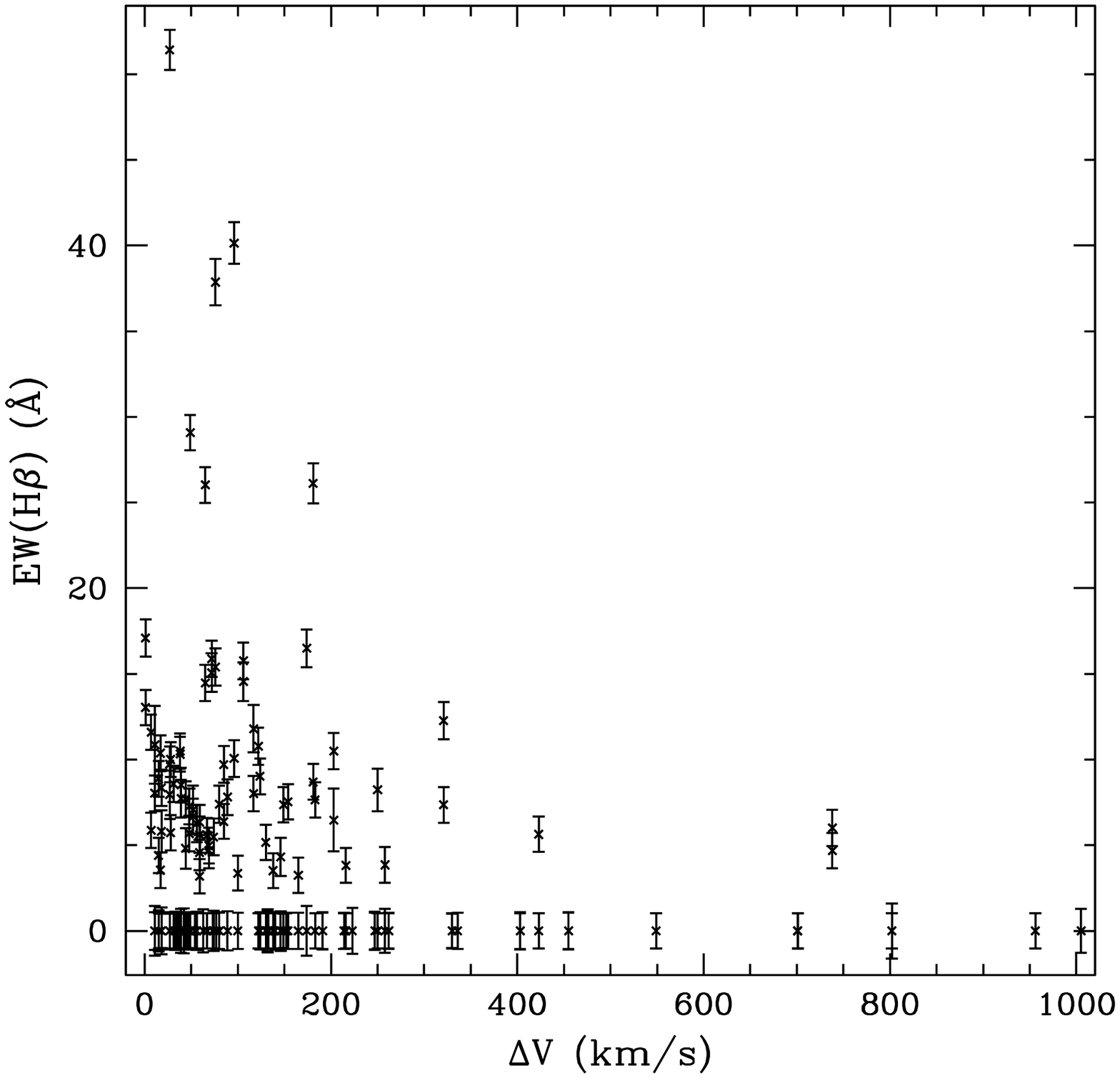}
\caption{Same as Fig.~5, except
that the ordinate is EW(H$\beta$).}
\label{fig:ol1}
\end{figure}

\begin{figure}
\plottwo{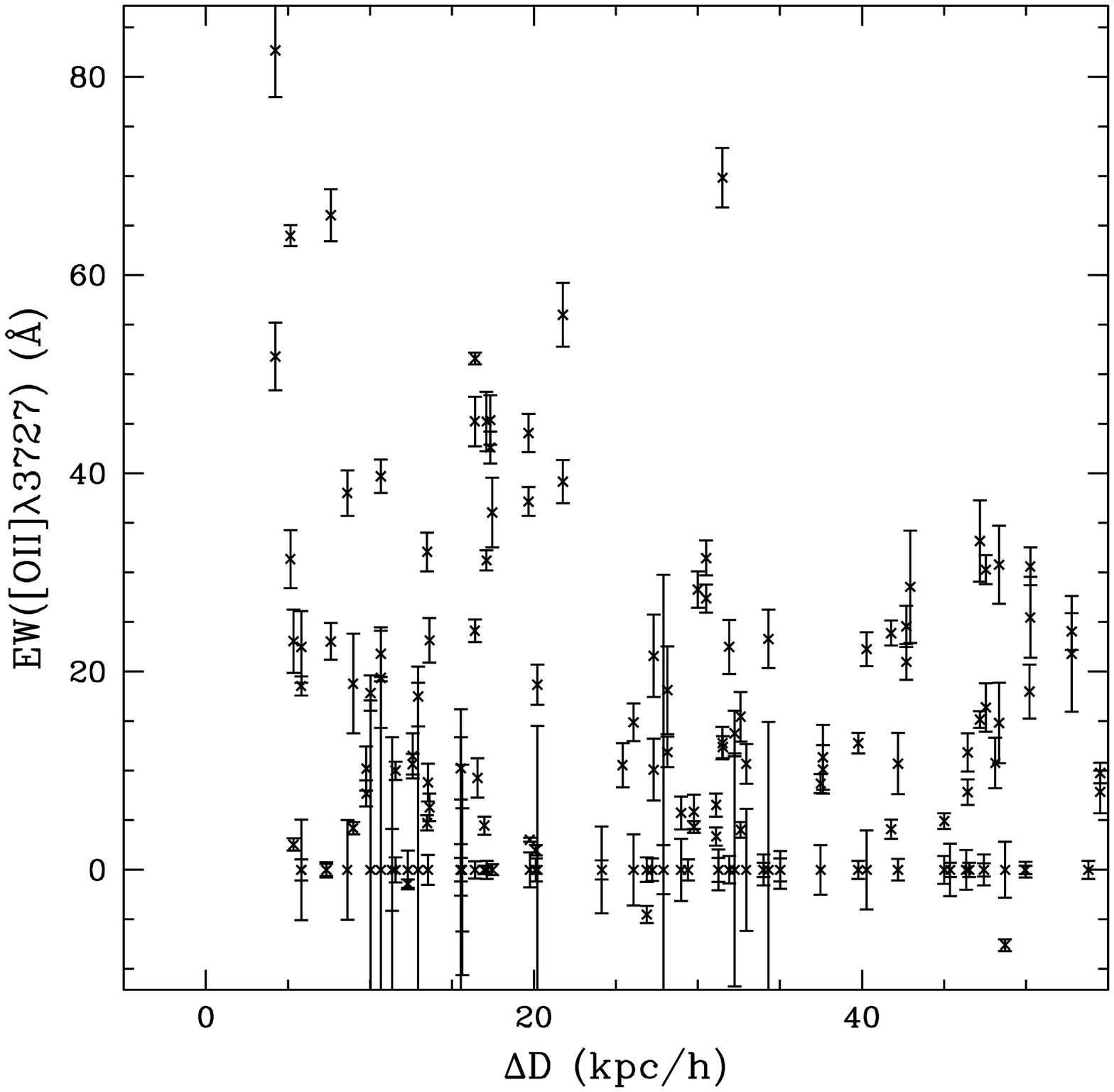}{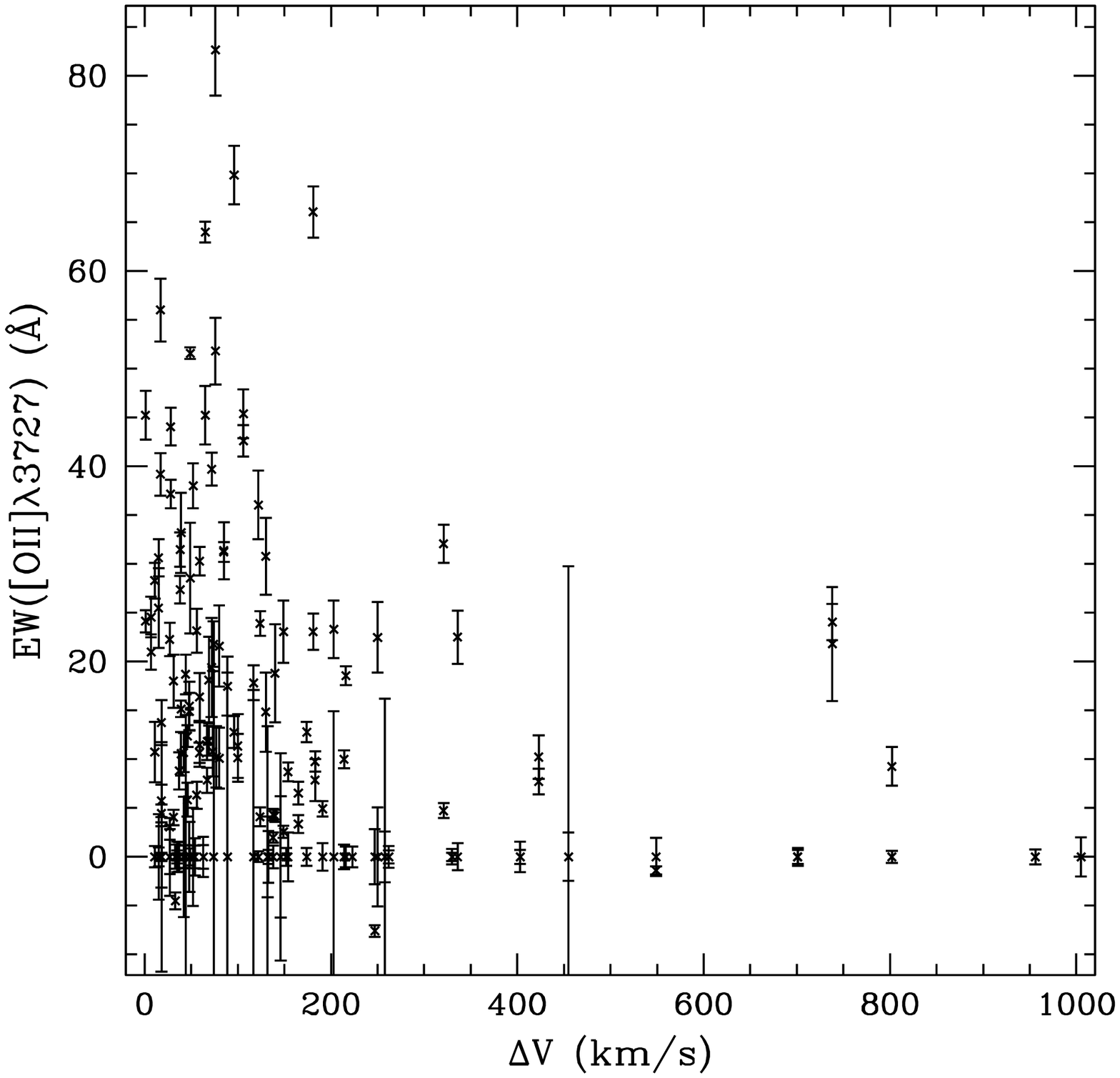}
\caption{Same as Fig.~5, except
that the ordinate is EW([OII]$\lambda3727$).}
\label{fig:ol2}
\end{figure}

\begin{figure}
\plottwo{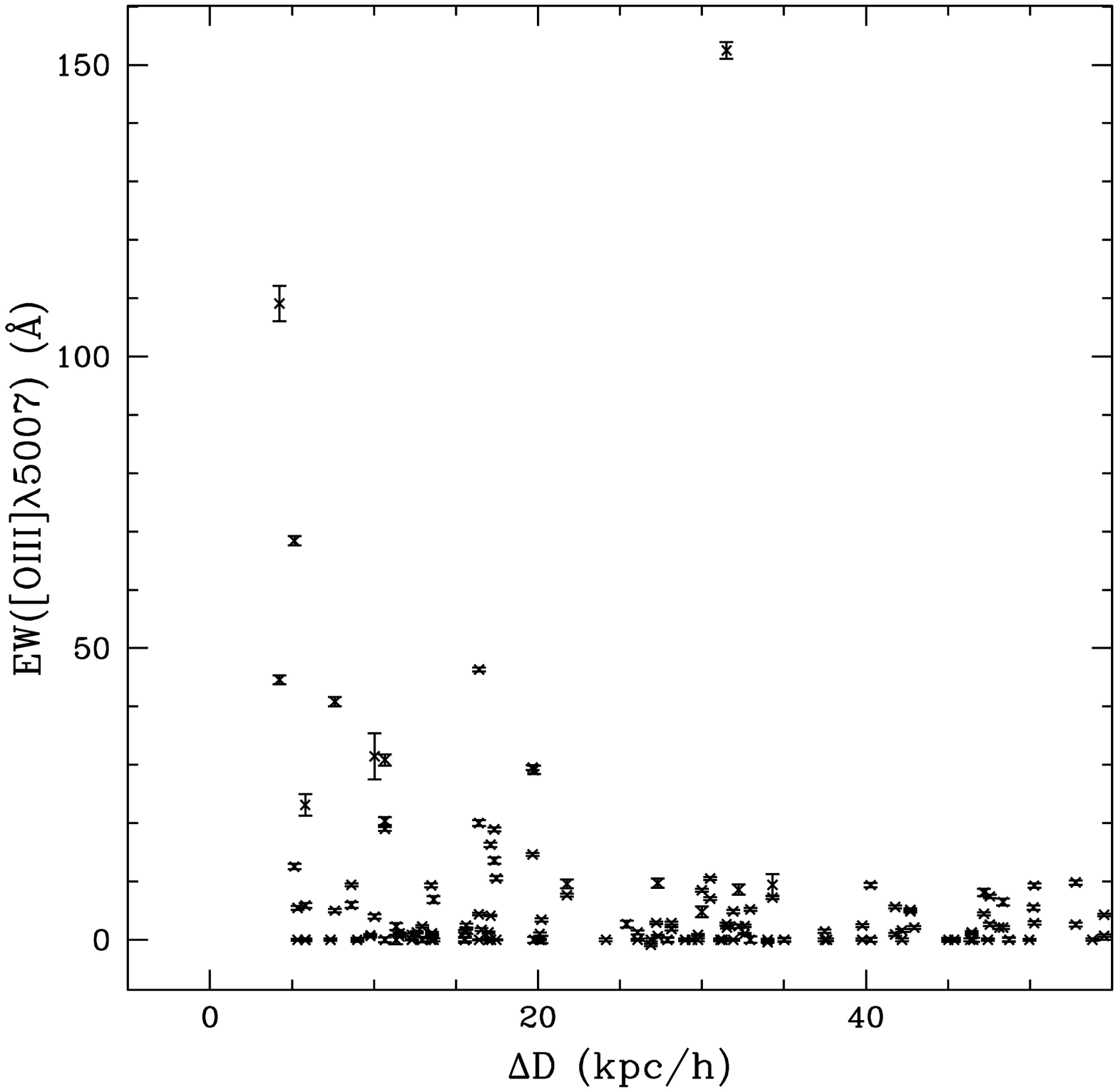}{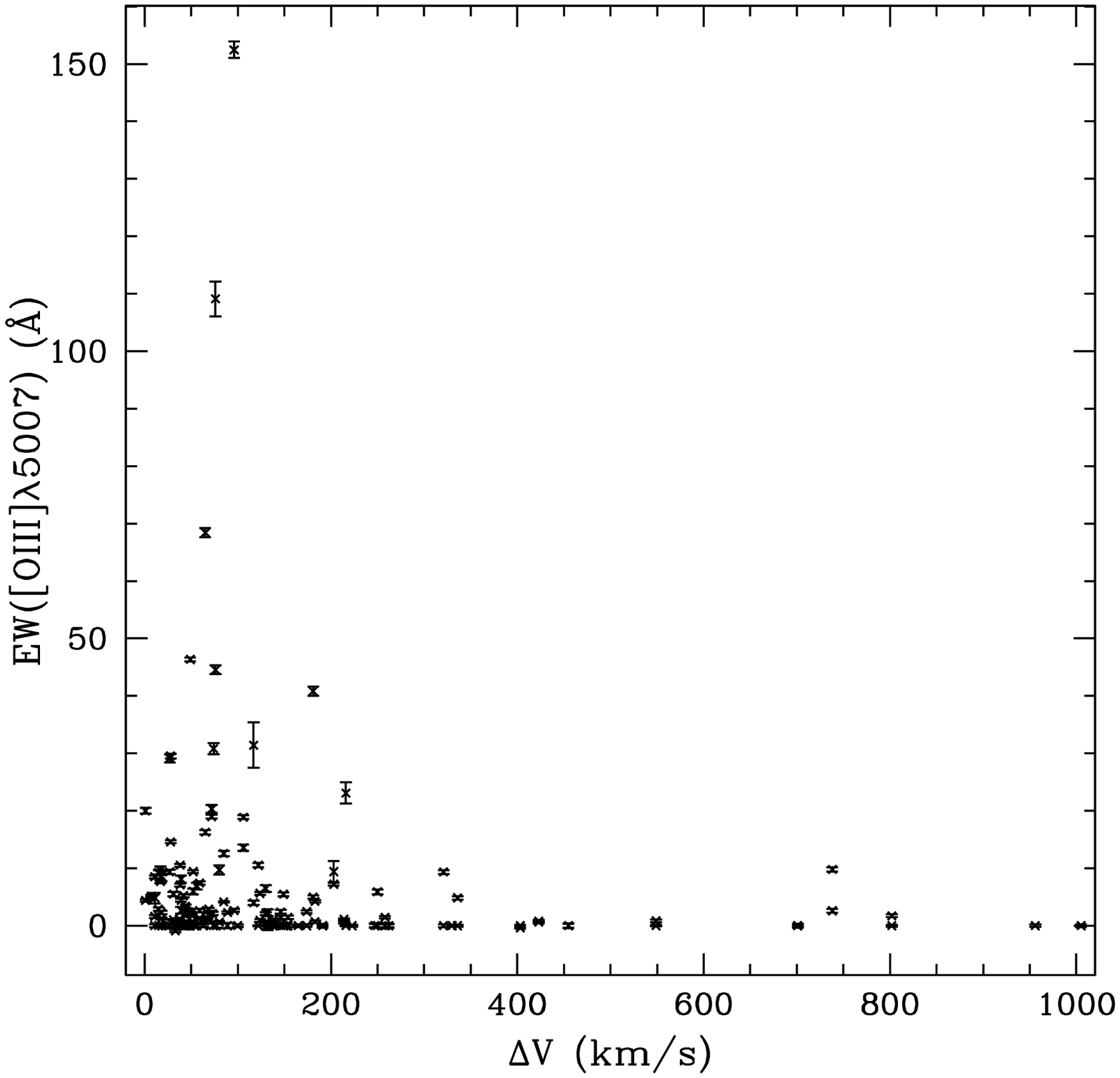}
\caption{Same as Fig.~5, except
that the ordinate is EW([OIII]$\lambda5007$).}
\label{fig:ol3}
\end{figure}

Other emission lines are more
accessible in high-$z$ systems.
Figs.~\ref{fig:ol1}~--~\ref{fig:ol3} show $\Delta D$ and
$\Delta V$ vs  equivalent widths for H$\beta$, [OII]($\lambda$3727)
and [OIII]($\lambda$5007), including only 
galaxies in environments where $\rho_{2.5} \leq 2.2$.  
We omit one galaxy in Fig.~\ref{fig:ol2} with no detection and 
huge error bars ($> 50$~\AA).  The trends we observe in
H$\alpha$ remain for each of these lines; Table~\ref{tab:ol1} 
shows the Spearman rank probabilities of no trend for all of 
these lines and for H$\alpha$. The [OIII] correlation with 
$\Delta D$ is the strongest among the collisionally-excited lines.  

As we note in Sec.~2, the CfA2North subsample is biased toward
galaxies with emission lines, although the line-emitting fraction of the
CfA2North subsample is unbiased with respect to the line-emitting
fraction of the CfA2South subsample.  The dearth of non-line-emitting
galaxies is unlikely to affect the correlation, because these
galaxies are distributed roughly evenly in $\Delta D$ in 
both the North and South. However, 
as a final check that the correlation we
detect is real, we also restrict our test to the unbiased
sample of galaxies with
${\rm EW}({\rm H}\alpha) > 0$ and $\rho_{2.5} \leq 2.2$.  This 
subsample includes only 106 galaxies, greatly reducing the ability
of the Spearman rank test to detect the correlation.  Nevertheless,
P$_{\rm SR} = 7.0 \times 10^{-2}$ and 
P$_{\rm SR} = 1.3 \times 10^{-2}$ for
$\Delta D$~--~EW(H$\alpha$) and  $\Delta D$~--~EW(HeI)
correlations, respectively.  The stronger HeI correlation 
is consistent with the association between HeI 
and more recent star formation.  The significance
of the trends for other lines and for $\Delta V$ is greatly
reduced. We record these statistics in Table~\ref{tab:ol1}.  

Any complete sample of galaxies in pairs undoubtedly contains 
galaxies which do not show this starbursting effect because (1) 
they are interlopers, (2) they have not yet reached the stage
at which the burst will begin,
or (3) they are not gas-rich.  Therefore, 
the correlations between emission line properties and
orbit parameters in complete samples of 
optically-selected pairs may be weak.  However, we conclude 
on the basis of rigorous tests that the emission line
properties of galaxies in pairs correlate with $\Delta D$, 
independent of any sampling biases we detect.  
The correlations are not the result of
large-scale environmental effects, such as gas stripping or the 
morphology-density relation.  We conclude that 
these correlations arise from local, tidal interactions.

The correlations in our data are completely 
model-independent. However, recent simulations suggest a 
physical basis for them (Sec.~5).
In the MH96 simulations, close orbital passes induce bursts of
star formation in galaxies with shallow central
potentials.  These models essentially predict that
emission-line properties of galaxies in pairs should depend on the 
local physics, including the parameters of the orbit.  
The correlations we find show that the local physics
has a substantial 
affect on the emission properties of galaxies in pairs. 

\section{Starburst Models of Orbiting Galaxies} 

Here, we show that the correlation between 
$\Delta D$ and EW(H$\alpha$) can plausibly arise when a 
close pass between the galaxies triggers a burst of star formation which 
continues as the galaxies move apart.  As the burst continues,
EW(H$\alpha$) decreases as the new population raises
the continuum around H$\alpha$.  Roughly, 
$\Delta D$ increases and EW(H$\alpha$) decreases after the close
pass, giving rise to the correlation.

We apply the starburst models of Leitherer et al. 
(1999; L99 hereafter; upgraded versions of the models
in Leitherer \& Heckman 1995).  In Sec.~7,
we use the L99 models and the EW(H$\alpha$) 
measurement in each galaxy
to estimate the time since a recent 
starburst began, $t_{\rm burst}$.  In Sec.~8, we 
discuss some consistency checks of the model, and in Sec.~9 we 
discuss sources of scatter and offsets
in the transformation from an EW(H$\alpha$)
to a $t_{\rm burst}$ or to a time since a close pass between
the galaxies, $t_{\rm pass}$.

\subsection{The Leitherer et al. Models}

L99 model a starburst by ``forming'' stars according to an
initial mass function (IMF) $\frac{\rm dN}{\rm dm} = C{\rm m}^{-\alpha}$,
where m is the stellar mass, $\alpha$ is the slope of the IMF,
and $C$ is a constant. 
They use models from the literature or construct their own models
for stellar evolution, stellar atmospheres, and nebular emission.
They compute spectral features, including EW(H$\alpha$),
as a function of time for both the ``instantaneous'' starburst case, 
with a timescale $\tau \lesssim 10^6$ years, and the continuous case, 
with star formation over $10^9$ years.  

The L99 models have a range of metallicity (0.05 Z$_{\sun}$, 
0.2 Z$_{\sun}$, 0.4 Z$_{\sun}$, Z$_{\sun}$, 2Z$_{\sun}$).  
The models also have a range of IMFs, including an approximate 
Salpeter (1955) IMF, with $\alpha=2.35$, 
from 1~M$_{\sun}$~--~100~M$_{\sun}$, a Salpeter IMF with a 
30~M$_{\sun}$ high-mass cutoff, and an approximate 
Miller-Scalo (1979) IMF with $\alpha=3.3$ from
1~M$_{\sun}$~--~100~M$_{\sun}$.  Hereafter, we refer to these 
IMFs as Salpeter, cutoff Salpeter, and Miller-Scalo, respectively.
In the continuous Miller-Scalo case, there are only very 
slight differences in EW(H$\alpha$) as a function of
time between the Leitherer \& Heckman (1995) model and the L99
model.

\subsection{Measuring the Time Since the Burst Began With H$\alpha$}

We use the model results from L99 to convert 
EW(H$\alpha$) for each galaxy in an environment 
with $\rho_{2.5} \leq 2.2$ to
an estimated time since the burst began, $t_{\rm burst}$.
Inclusion of the points with $\rho_{2.5} > 2.2$ has
no qualitative effects on the results.
EW(H$\alpha$) decreases monotonically with time.
Thus, the Spearman rank test significance of the 
$\Delta D$~--~$t_{\rm burst}$ correlation under these assumptions
should be the same as that of the $\Delta D$~--~EW(H$\alpha$)
correlation (Sec.~6).
We begin applying the models assuming that the measured
EW(H$\alpha$) depends on timescale according
to the L99 models --- that no significant older population 
contributes to the continuum around H$\alpha$.  Because of
possible contributions from old populations, EW(H$\alpha$) is 
a lower limit to the EW(H$\alpha$) due to the burst alone, 
and the $t_{\rm burst}$ we compute is an upper limit to the true
time since the burst, in the context of the L99 models.

Barton et al. (1999b) use photometry of
190 galaxies in the spectroscopic sample
and longslit spectra of 84 spirals and S0s to
investigate the possible effects from an
old stellar population.  They apply the Leitherer et al. (1999)
starburst models to measure s$_{\rm R}$, the fractional $R$-band 
flux of the new burst, and $t_{\rm burst}$, the elapsed time since 
the burst began.  With their assumptions about the color of the old 
stellar population and the reddening correction
(based on the observed Balmer decrement), 
the Leitherer et al. (1999) Miller-Scalo IMF models 
imply large burst strengths ($0.2 \lesssim {\rm s}_{\rm R} \lesssim 1$) 
in the centers of some of the galaxies.
The burst strength, s$_{\rm R}$, may depend
on at least one internal parameter of the galaxies, the velocity width.  
The interaction-related parameters appear to
determine $t_{\rm burst}$; the measured values of 
$t_{\rm burst}$ correlate significantly with $\Delta D$, the projected
pair separation on the sky.
Thus, although their conclusions are somewhat sensitive to assumptions
about reddening, the IMF of the new burst, and the color of 
the old stellar population, Barton et al. (1999b) show
that in many galaxies, the burst population may constitute a significant
fraction of the $R$-band flux ($\gtrsim 20\%$, and up to
$\sim$100\%).  For bursts of this strength, contributions from
the old stellar population have only a small effect on the 
measured $t_{\rm burst}$. Thus the approach we take here of ignoring
the old population should give a reasonoble indication of the relationship
between $t_{\rm burst}$ and the kinematic parameters of the
interaction.

We apply the two star formation models of L99:
the instantaneous burst and the continuous burst. The star
formation rate probably depends on the interaction strength and the 
galaxy potential (MH96); however, Leitherer \& Heckman (1995) show
that these two cases successfully bracket the properties of
intermediate cases.  Later, we apply
consistency checks and discuss complications which affect
the interpretation of the results.

\begin{figure}
\plottwo{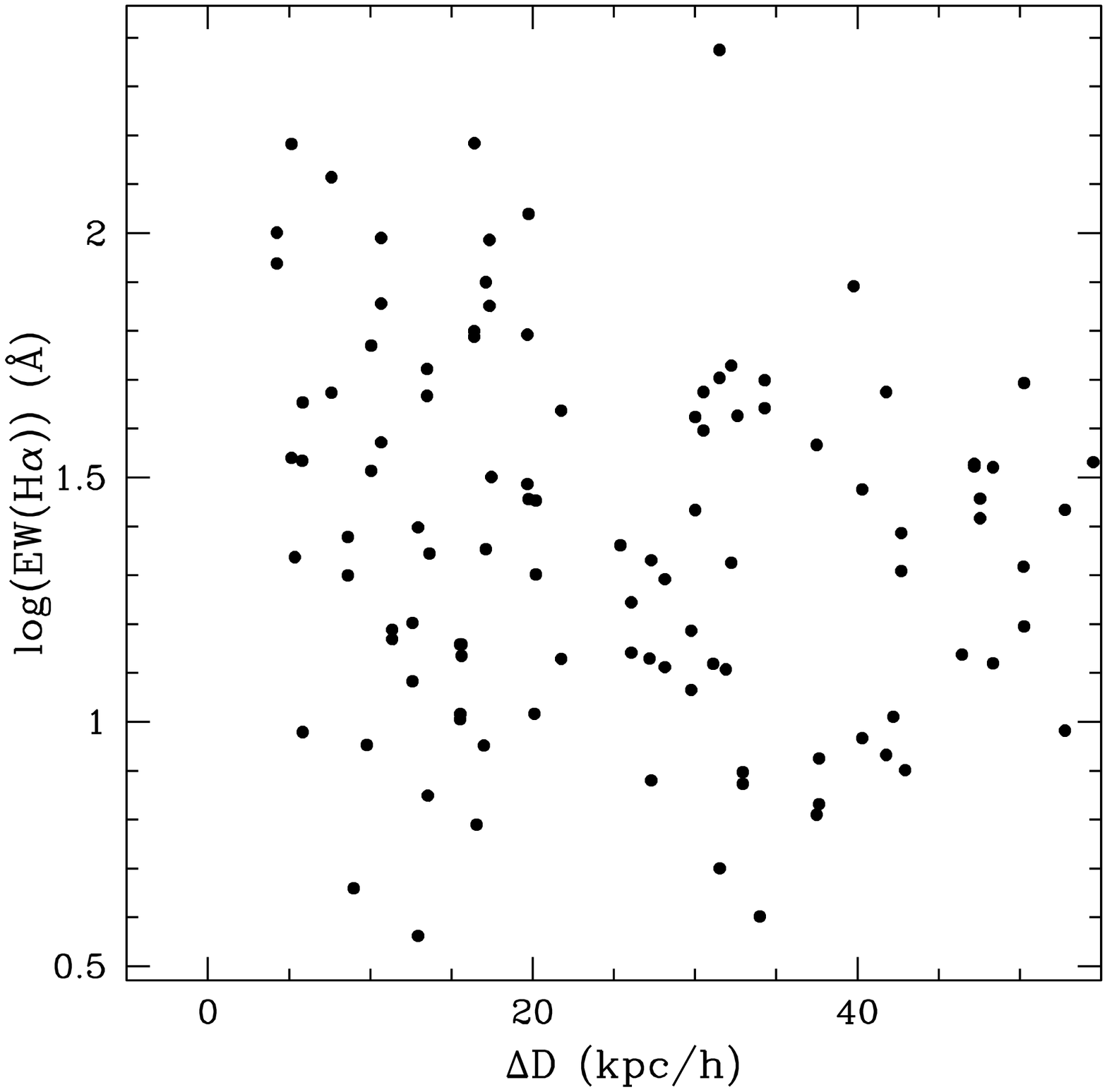}{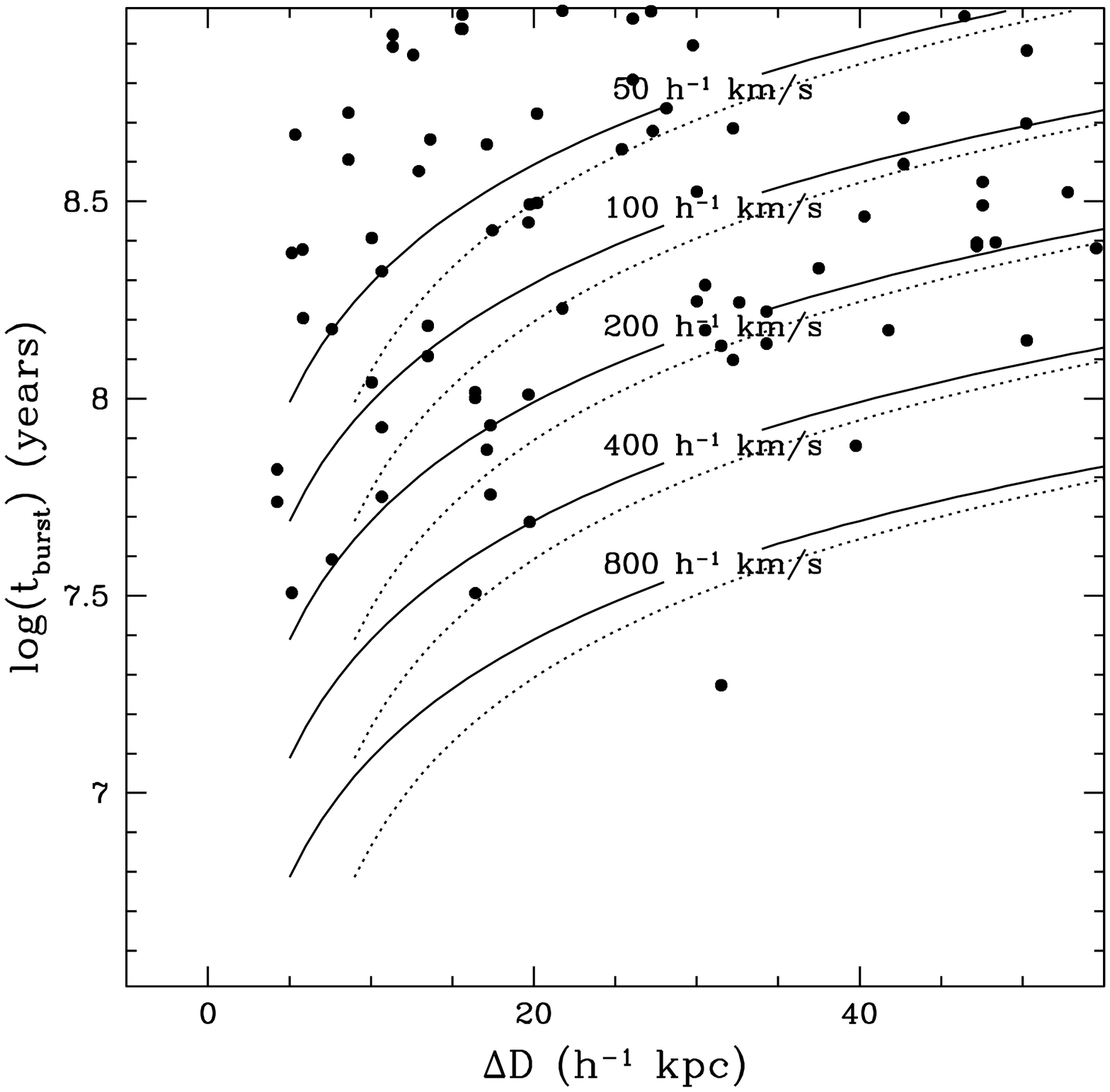}
\caption{(a) $\Delta D$ vs. log(EW(H$\alpha$)) for galaxies
with EW(H$\alpha$)$\gtrsim 10$\AA, and (b) the corresponding 
log($t_{\rm burst}$) from the L99 model with continuous star
formation, Z=Z$_{\sun}$ and a Miller-Scalo IMF.  
The solid contours are lines of constant average velocity in the plane of
the sky; the dotted contours are offset by 4~h$^{-1}$~ kpc.  
Only points with
$t_{\rm burst} < 10^9$~years appear in (b).}
\label{fig:lh1}
\end{figure}

Fig.~\ref{fig:lh1} shows $t_{\rm burst}$ computed with 
a Miller-Scalo IMF with Z$=$Z$_{\sun}$ and continuous star formation. 
Fig.~\ref{fig:lh1}a shows $\log({\rm EW}({\rm H}\alpha))$
as a function of $\Delta D$ for the line-emitting
galaxies.  We correct EW(H$\alpha$) for Balmer absorption 
because L99 do not account for it in their computation. 
Fig.~\ref{fig:lh1}b shows the log of the corresponding time in years 
since initiation of the burst; we convert each equivalent width to
a $t_{\rm burst}$ using 
simple linear interpolation of the L99 points. In this and all
similar plots, we exclude galaxies with $t_{\rm burst} > 10^9$ years:
the L99 models do not extend past 1 Gyr. 

This correlation between $\Delta D$ and $t_{\rm burst}$ depends
only on our assumptions about the behavior of the burst 
population and the contribution of other populations to the
spectrum; it is independent of the predictions of the n-body
models.  However, if it does indicate that pairs with larger
$\Delta D$ have older bursts, what were the (spatial) properties of the 
galaxies when the bursts were younger?  Presumably, they were similar
to the youngest bursts now: they had small pair separations ($\Delta D$).  
Thus, the interpretation based on the MH96 models
follows naturally: bursts are initiated at small $\Delta D$ and
age as the galaxies move apart.

\begin{figure}
\centerline{\epsfysize=7in%
\epsffile{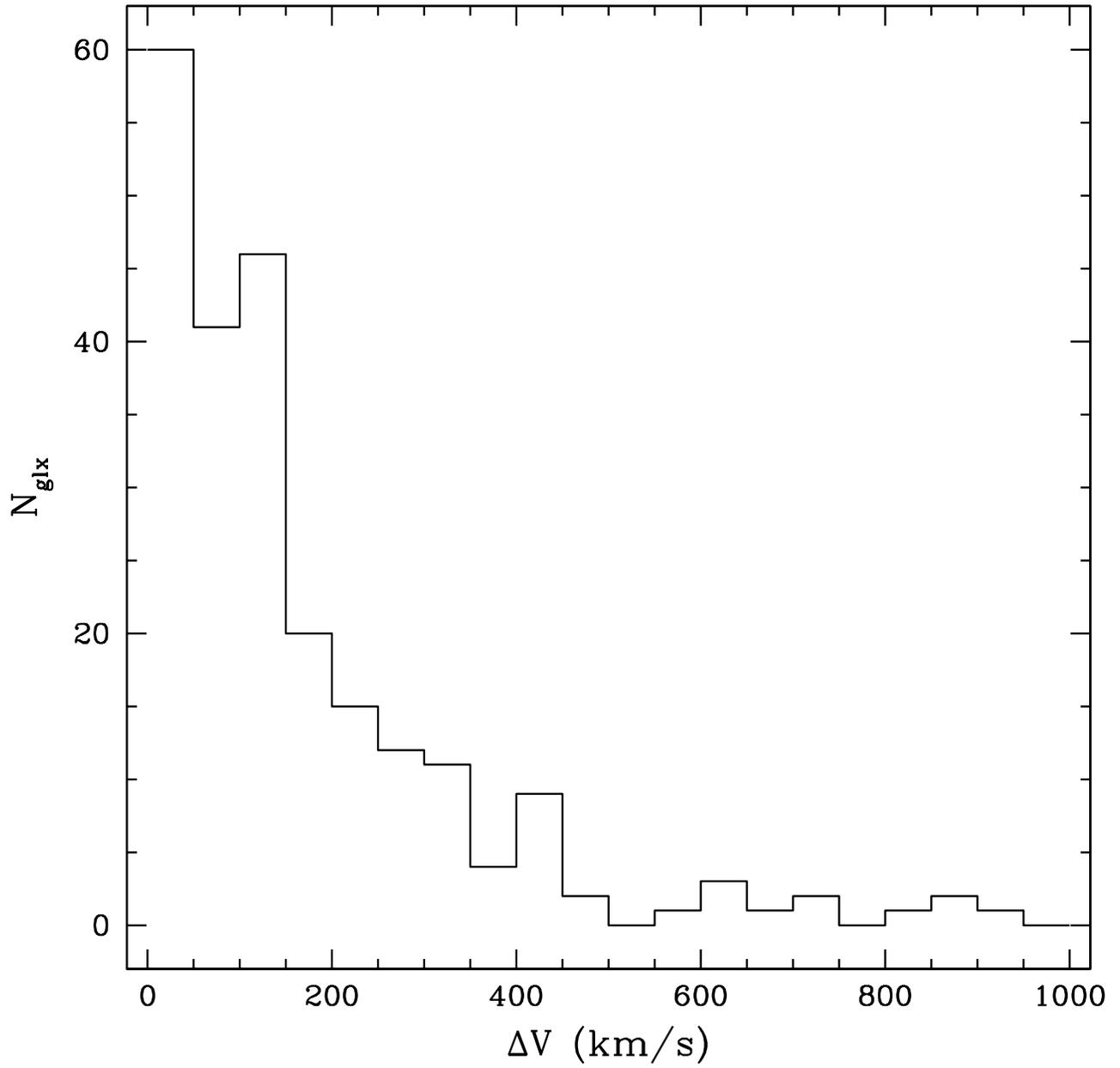}}
\caption{Line-of-sight velocity differences
for galaxies in the sample with significant H$\alpha$ emission.}
\label{fig:lhv}
\end{figure}

If interaction between the two galaxies in each pair did give rise to a
starburst which resulted in the observed EW(H$\alpha$),
the current separation of the galaxies and the 
elapsed $t_{\rm burst}$ should provide information about
the average velocity of the pair in the sky plane.  
Thus, in Fig.~\ref{fig:lh1}b we plot contours of 
``average velocity'', $\bar{v}_{\bot}$, in the plane of the sky, 
where $\bar{v}_{\bot}=\Delta D/$$t_{\rm burst}$.  
Although we cannot measure either component of this 
velocity for independent comparison,
the size of our sample allows statistical estimation of the
appropriate velocity range.  The range is 
$\sim \sqrt{2} \times (50$~--~$200$~km/s), assuming no 
preferred orientation
of orbits, based on the distribution of the line-of-sight 
velocity separation for
pairs with H$\alpha$ (Fig.~\ref{fig:lhv}).  In the
next section, we explore the expected distribution in detail.

If the starburst began at exactly the time of a very close pass, 
the time at which $\Delta D \approx 0$ for each pair, 
$\bar{v}_{\bot}$ is the average relative velocity of the pair in the 
plane of the sky since the close pass.  For nonzero impact 
parameter ($\Delta D = b \neq 0$ at closest approach), 
the $\bar{v}_{\bot}$ contours should be shifted to the
right by the value of $b$.  The dotted contours are $\bar{v}_{\bot}$
shifted by $b = 4 {\rm\ h}^{-1}$~kpc, approximately the 
maximum 3-dimensional $b$, and hence the maximum 2-dimensional $b$, 
such that the 3-dimensional turnaround radius of the orbit is 
$\leq 50 {\rm\ h}^{-1}$~kpc according to our rough orbit models 
(see Sec. 7.3).  Thus, in the context of our orbit model, 
the range between the solid and dotted contours roughly represents 
the expected scatter due to nonzero $\Delta D$ at closest approach.

The essence of this analysis is the comparison of the starburst
timescale with the dynamical timescale in these systems.  This 
approach has been successful in the past, when applied to small 
numbers of mergers with burst ages estimated from the star cluster
populations (Whitmore et al. 1997 and references therein).  
Fig.~\ref{fig:lh1}b shows that the scales match appropriately;
the points with the largest EW(H$\alpha$) in 
Fig.~\ref{fig:lh1}b have relative velocities in the plane of
the sky (corresponding to the contours) which fall very nearly in the
proper range, $\lesssim 50$~km/s to 
$\sim 400$~km/s, roughly 
as expected from the {\it independent} line-of-sight relative
velocities (Fig.~\ref{fig:lhv}).  
The plot shows one outlier, the starburst nucleus UGC~2103, 
at $\geq 800$~km/s.
The shape of the lower edge of the locus of the points with
large EW(H$\alpha$) is a good match to an approximately constant 
velocity of $\sim 400$~km/s.  Next,
we explore the expected distribution of points in this
$\Delta D$-$t_{\rm burst}$ plane based on simple orbit models.
In Sec.~7.4, we apply the other variants of the L99 models.

\subsection{Comparing the Data to Orbit Models}

The expected distribution of $\bar{v}_{\bot}$ for a sample of
orbiting pairs depends on the population of the orbits and on
the effects of dynamical friction (Turner 1976a; 1976b;
Bartlett \& Charlton 1995).
Fig.~\ref{fig:or1} shows the distribution in the 
$\Delta D$~--~$t_{\rm burst}$ plane for a population of galaxies 
just after the first pass
for orbits with zero initial orbital energy and with the 
characteristics listed in the figure caption.
In each figure, we include 
19 orbits with a range of impact parameters 
at first pass, $b = 2$~--~24 h$^{-1}$ kpc in increments of ~1.3 h$^{-1}$ 
kpc (we assume H$_0$~=~65~km/s/Mpc to convert from kpc in the
models to the units of observation, h$^{-1}$~kpc).  We weight orbits
by $b_{\rm Kep}$, the value of $b$ in the absence of dynamical
friction, to account for the larger probability
of a larger impact parameter.  We compute the orbits using a rough 
implementation of the Chandrasekhar dynamical friction formula 
(Chandrasekhar 1943; Binney \& Tremaine 1987) for isothermal 
spheres.  We tune the calculation for a very rough match to orbits 
based on the self-consistent n-body models of Barton et al. (1999a).  

Other orbit families with varied amounts of initial orbital energy 
and/or varied galaxy masses occupy loci on Fig.~\ref{fig:or1} 
with similar shapes, but the quantitative locations of the edges 
of the point distribution change substantially.  
Without {\it a priori} knowledge of galaxy masses and initial orbital
energies, as well as detailed treatment of dynamical friction,
we cannot construct an accurate set of orbit models 
to match to the data precisely.
Qualitatively, however, the models show that a population of orbits
fills in a region with a shape similar to the one occupied by
the data in Fig.~\ref{fig:lh1}b.

\subsection{Other L99 Models}

\begin{figure}
\centerline{\epsfysize=7in%
\epsffile{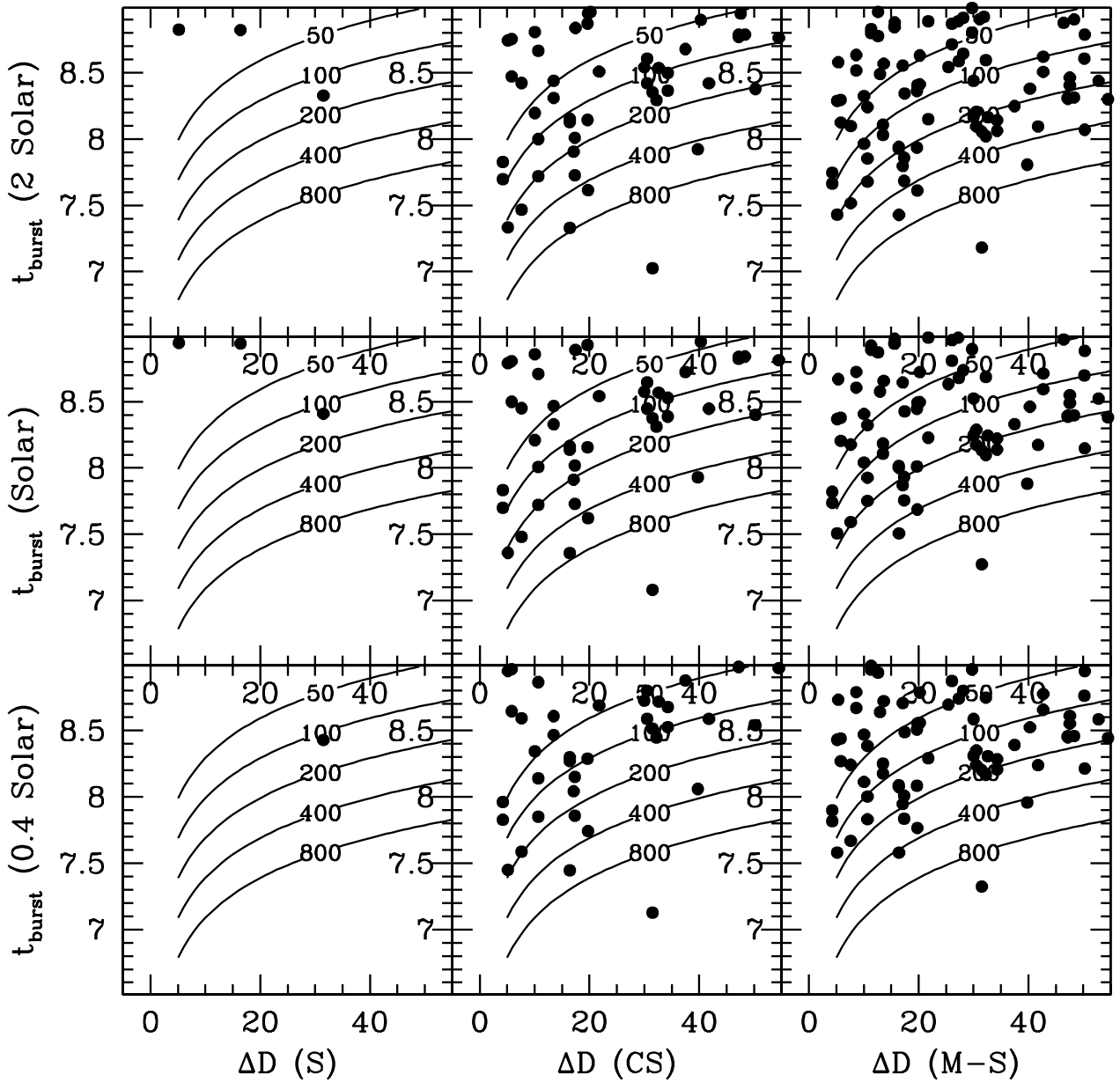}}
\caption{Same as Fig.~10b for a variety of L99 IMFs and metallicities,
in the continuous star formation case.  
The solid contours are lines of constant average velocity 
in the plane of the sky,in units of h$^{-1}$ km/s; 
we omit the dashed lines of Fig.~10b for simplicity.
S, CS, and M-S refer to the Salpeter, Cutoff Salpeter, and
Miller-Scalo IMFs, respectively.}
\label{fig:lh2}
\end{figure}

\begin{figure}
\centerline{\epsfxsize=7in%
\epsffile{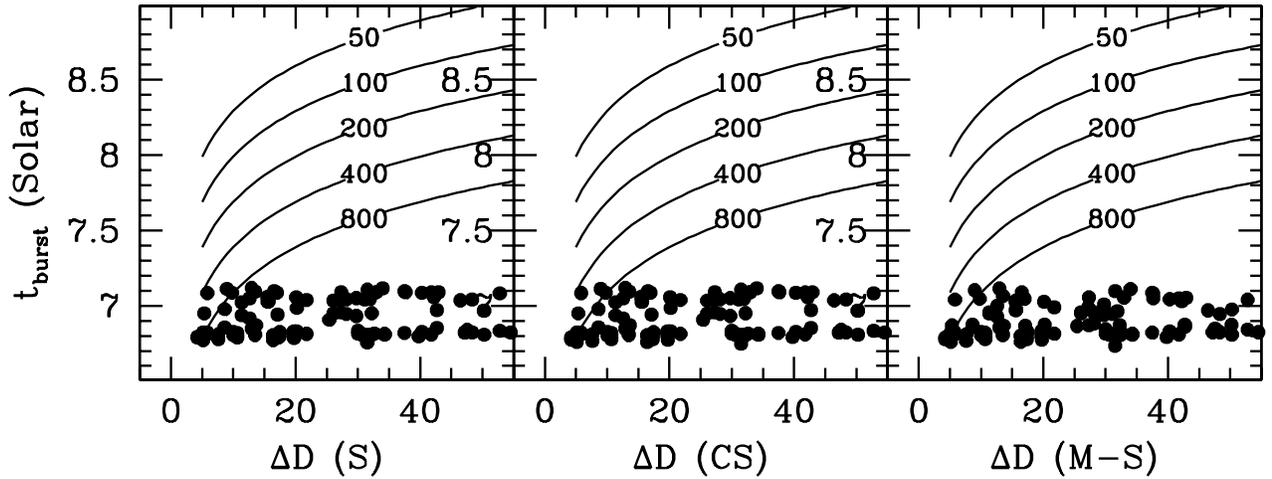}}
\caption{Same as Fig.~10b for a variety of L99 IMFs and solar metallicity,
in the instantaneous star formation case.  
The solid contours are lines of constant average velocity 
in the plane of the sky,in units of h$^{-1}$ km/s.
The plots appear very 
similar for $Z=0.4 Z_{\sun}$ and $Z = 2 Z_{\sun}$.
S, CS, and M-S refer to the Salpeter, Cutoff Salpeter, and
Miller-Scalo IMFs, respectively.}
\label{fig:lh3}
\end{figure}

Figs.~\ref{fig:lh2} and \ref{fig:lh3} show $t_{\rm burst}$ estimates 
for the L99 models with Z=0.4 Z$_{\sun}$, Z=Z$_{\sun}$
and 2 Z$_{\sun}$ in the continuous and 
instantaneous cases, respectively.  The Miller-Scalo and 
cutoff Salpeter IMFs match better than the Salpeter IMF. The
Salpeter IMF points mostly fall above the graph --- the Salpeter
IMF predicts values of EW(H$\alpha$) much larger than
those in our sample, suggesting that massive stars are overabundant
in the Salpeter IMF.  For the instantaneous models, the required
velocities in the sky plane are generally $> 1000$~km/s,
much too large.  

\begin{figure}
\centerline{\epsfysize=7in%
\epsffile{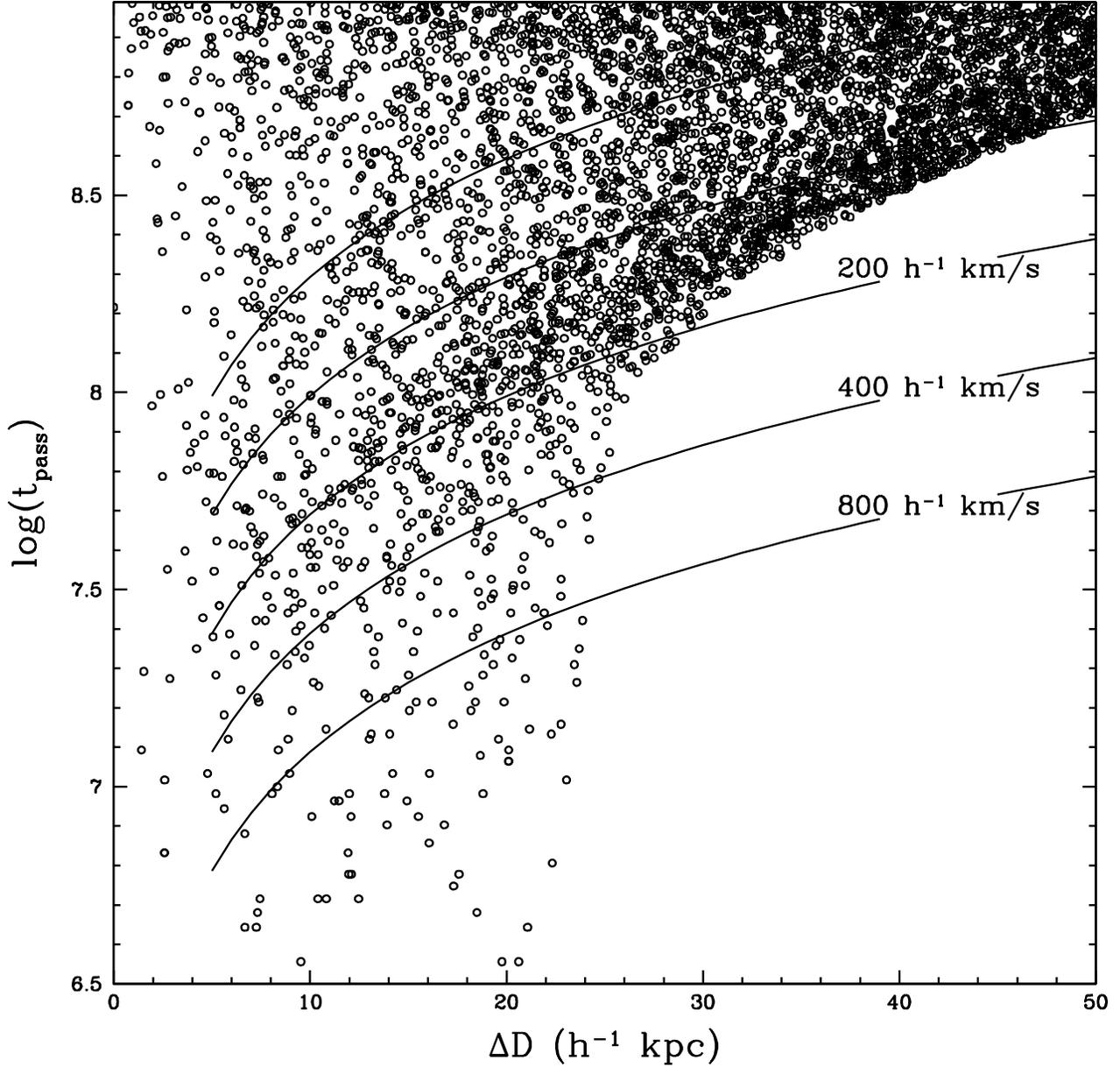}}
\caption{Orbit Models: $\Delta D$ vs. the log of the 
time since the most recent pass in years, 
log($t_{\rm pass}$), for a set of orbits with zero
initial orbital energy, E$_{0} = 0$.  The impact
parameters are spaced uniformly from 2 -- 24 h$^{-1}$ kpc, assuming
H$_{0} = 65$ km/s/Mpc, and orbits are weighted by $b$ to account for
the higher probability of a larger impact parameter.  
Points are at random times, viewed from
random angles.  The galaxy masses are M$_{\rm gal} = 5.8 \times 10^{11}$ 
M$_{\sun}$. The solid contours are lines of constant average velocity 
in the plane of the sky, for 50, 100, 200, 400 and 800 h$^{-1}$ km/s.}
\label{fig:or1}
\end{figure}

\begin{figure}
\centerline{\epsfysize=7in%
\epsffile{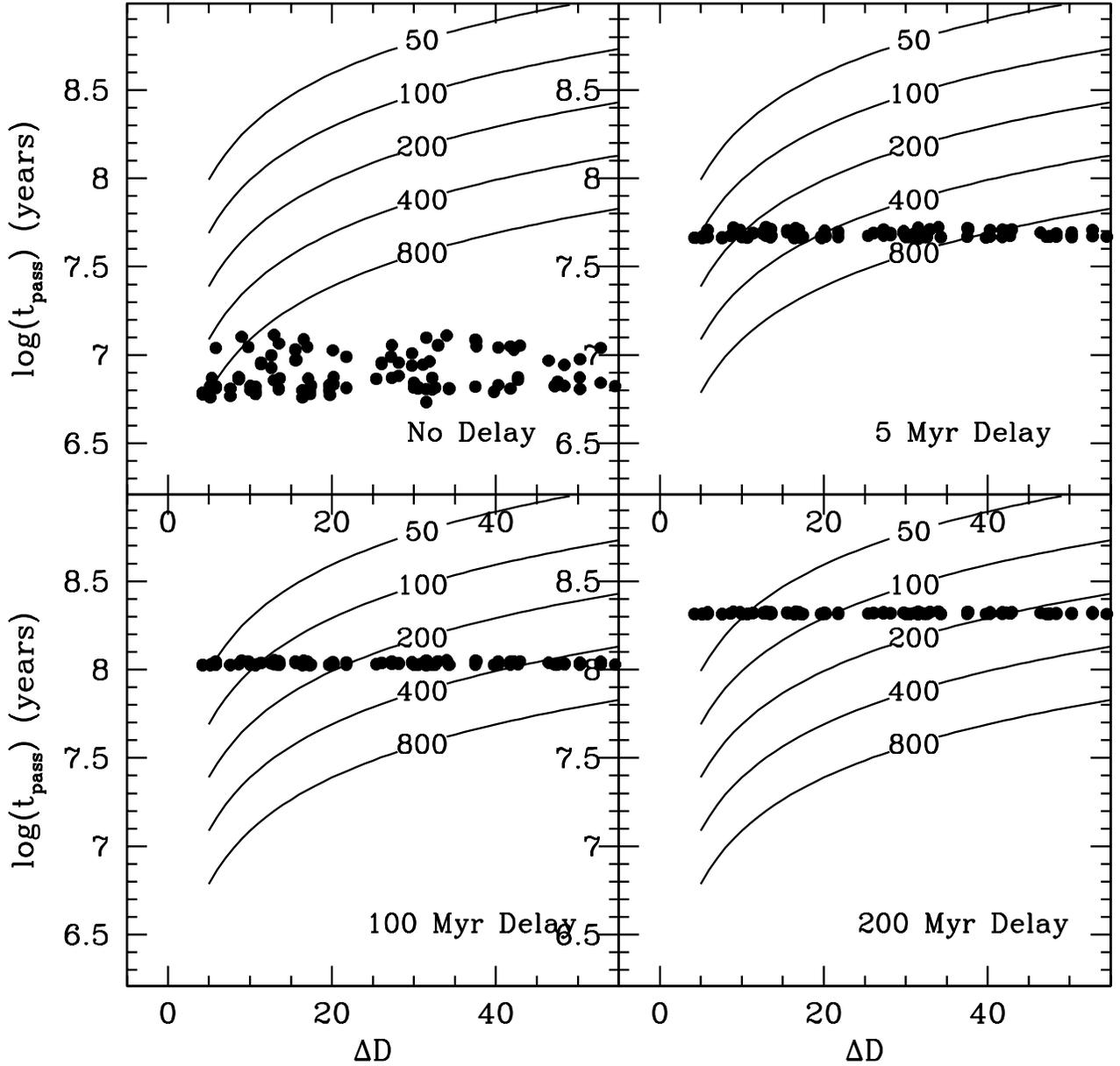}}
\caption{$\Delta D$ vs. log($t_{\rm pass}$ )
from the L99 model with instantaneous star
formation, Z=Z$_{\sun}$ and a Miller-Scalo IMF, for different
values of ${\rm t}_{\rm delay}={\rm t}_{\rm pass}-{\rm t}_
{\rm burst}$.  Only points with
$t_{\rm pass} < 10^9$~years appear.
The solid contours are lines of constant average velocity 
in the plane of the sky,in units of h$^{-1}$ km/s.}
\label{fig:lh4}
\end{figure}

However, direct comparison between Figs.~\ref{fig:lh2}~--~\ref{fig:lh3}
and Fig.~\ref{fig:or1} assumes that the
starburst began exactly at the time of the close pass, or
$t_{\rm burst} = t_{\rm pass}$.  In 
the MH96 models, the burst is delayed until $\sim$100 Myr
{\it after} the pass.  
Fig.~\ref{fig:lh4} shows times computed with an 
instantaneous model ($\alpha = 3.3$ IMF with Z$=$Z$_{\sun}$),
including delays in the starburst of $t_{\rm delay} =$
$t_{\rm pass}$~-~$t_{\rm burst} = $ 0, 5 Myr, 40 Myr and 200 Myr. 
The y-axis in these plots is now the time since the close
pass, $t_{\rm pass}$, not the time since the starburst; 
thus, the same $\bar{v}_{\bot}$ contours still apply.  A delay in the burst
can boost the points into the appropriate range of $\bar{v}_{\bot}$, but
the model no longer explains the observed EW(H$\alpha$) distribution.

In the instantaneous models, a large range in
EW(H$\alpha$) corresponds to only a small range of $t_{\rm burst}$.
Thus, if these instantaneous models were correct, most of the 
observed equivalent widths would correspond to $t_{\rm burst}
 \approx 10^7$~years, and would represent
only a brief snapshot in the interaction process.  If the 
instantaneous models were correct, something entirely different
would be responsible for the anti-correlation of $\Delta D$ and
EW(H$\alpha$), such as an unlikely correlation between 
$t_{\rm delay}$ and $\Delta D$ that is fine-tuned to
time differences of 10$^7$ years.
We conclude that, barring significant problems with the
conversion from EW(H$\alpha$) to $t_{\rm burst}$, 
the instantaneous models are excluded by the data ---
starbursts due to interaction must be extended in
time ($\tau \gg 10^6$~years and probably $\tau \gtrsim 10^8$~years).

\begin{figure}
\centerline{\epsfysize=7in%
\epsffile{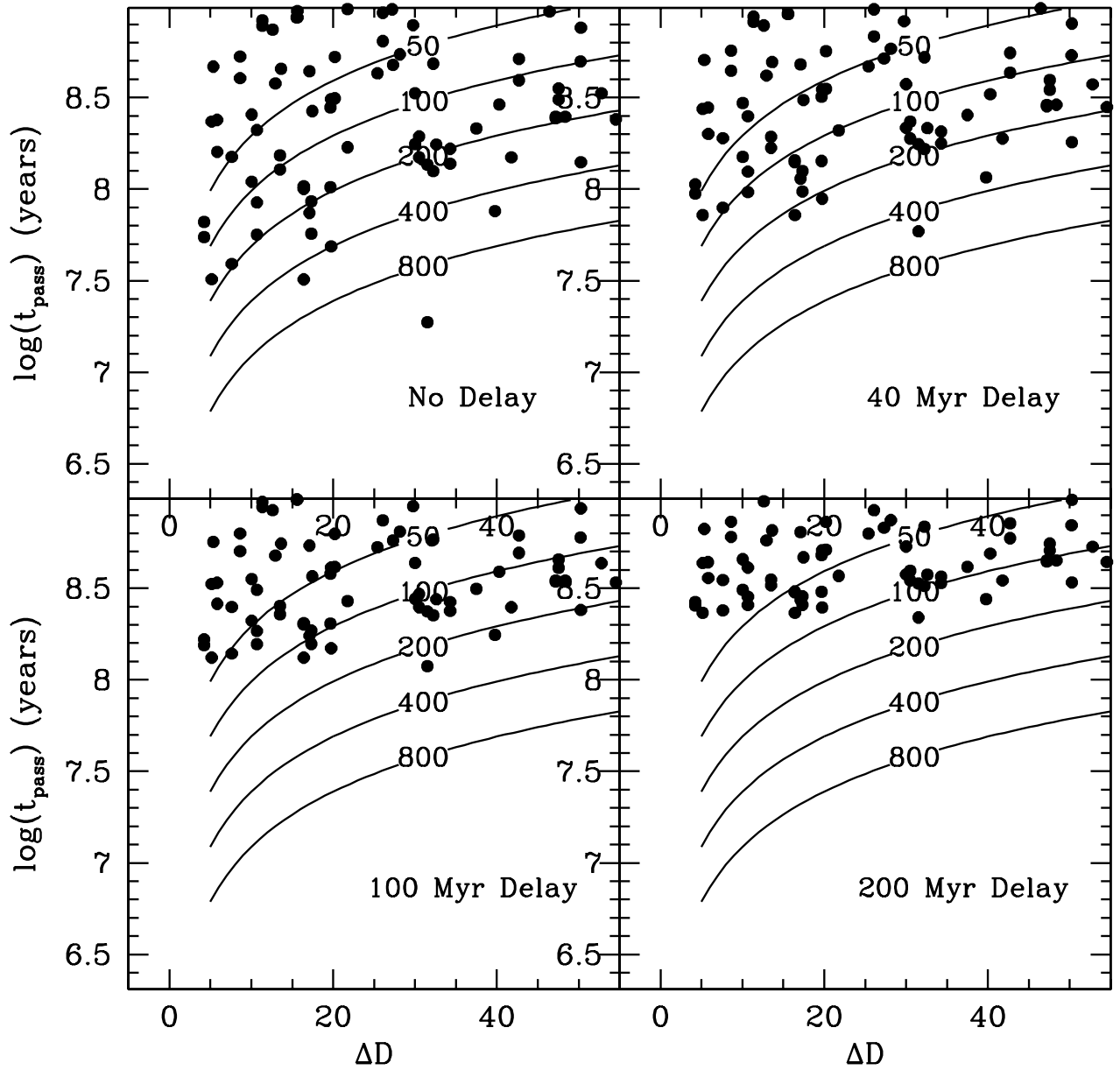}}
\caption{Same as Fig.~15 for continuous star formation.}
\label{fig:lh5}
\end{figure}

Because delays between the close pass and the burst may be
a generic feature of tidally-triggered starbursts, 
we explore delays in the continuous case as well.
In Fig.~\ref{fig:lh5} we plot $t_{\rm pass}$ for
$t_{\rm delay}$ = 0, 5 Myr, 40 Myr, and 200 Myr 
in the continuous models.  Large delays drastically 
change the locus of the points in the 
$\Delta D$~--~$t_{\rm pass}$ plane derived from the models.
The proper way to compare Fig.~\ref{fig:lh5} to 
Fig.~\ref{fig:or1} is to ignore the
points in Fig.~\ref{fig:or1} with
$t_{\rm pass} < $$t_{\rm delay}$; these points correspond to 
pre-burst galaxies where the EW(H$\alpha$) does not yet reflect
the close pass.  In this picture, the observed distribution 
allows only a moderate delay.  Similarly, the data do not rule out bursts 
initiated a very short time before closest approach.

The larger delays, with $t_{\rm delay} \gtrsim 50$~Myr, change
the shape of the distribution until it no longer tracks the
shape of the constant-velocity contours,
and the model no longer explains the correlation 
between EW(H$\alpha$) and $\Delta D$.  If our interpretation 
is correct, the data admit only the Miller-Scalo and cutoff 
Salpeter IMFs with continuous star formation.

\section{Consistency Checks for the Model}

Secs.~5 -- 7 suggest a simple interpretation of our
pair data that agrees with qualitative 
results from n-body/SPH simulations
(MH96), starburst models (L99), and simple orbit models.  
In Sec.~7 we show that timescales based on the several
L99 models match the dynamical timescales for a family of orbits,
explaining the observed $\Delta D$~--~EW(H$\alpha$) correlation.
This picture has implications for other aspects of the data which we explore
here, including
the $\Delta V$ distribution as a function of $t_{\rm pass}$, 
the strength of Helium lines, and the presence of Balmer absorption.

\subsection{Comparing to the $\Delta V$ Distribution} 

\begin{figure}
\centerline{\epsfysize=7in%
\epsffile{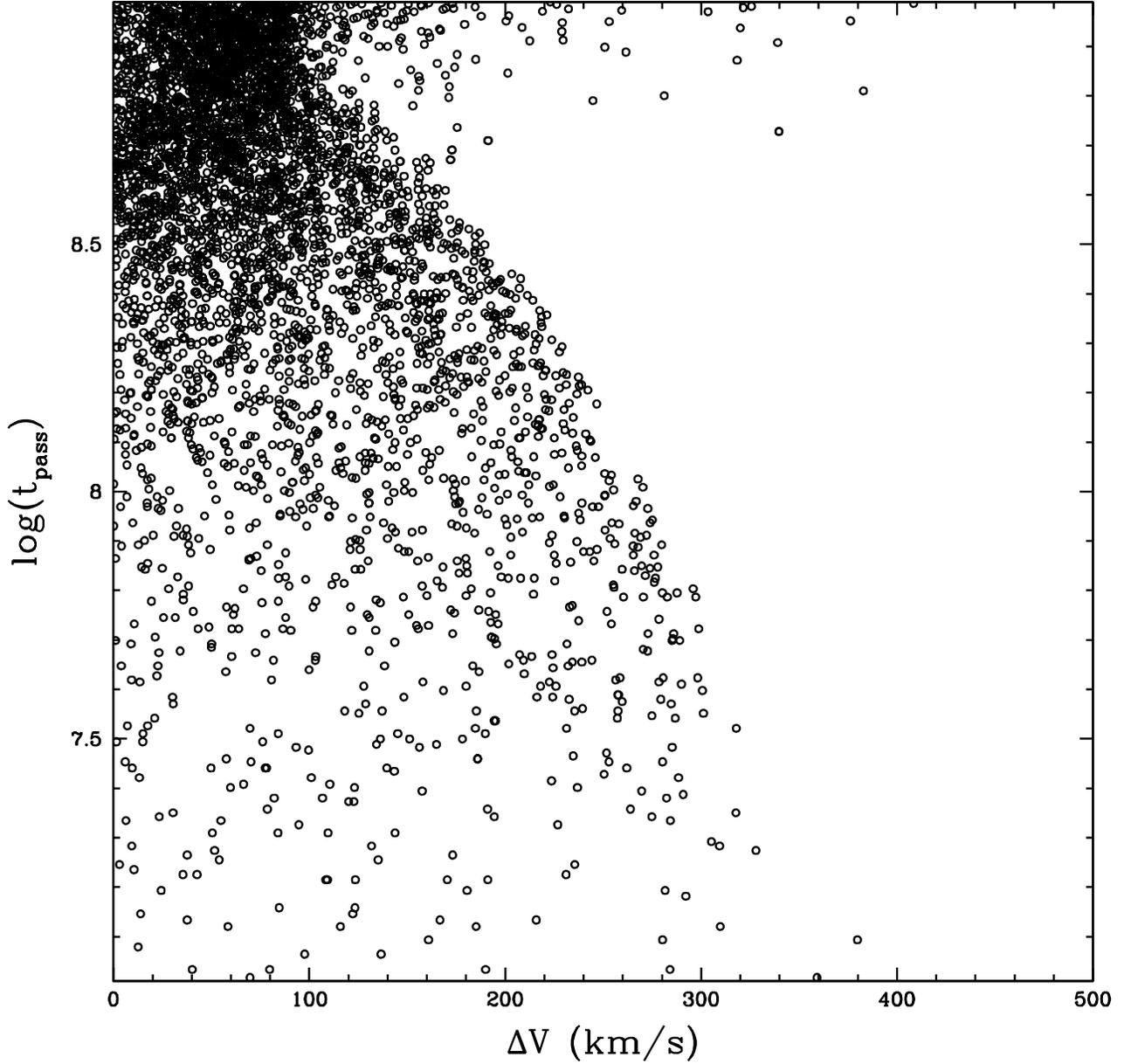}}
\caption{Orbit Models: $\Delta V$ vs. the log of the 
time since the closest pass in years, 
log($t_{\rm pass}$). 
The impact parameters are spaced uniformly 
from 2 -- 24 h$^{-1}$ kpc, assuming
H$_{0} = 65$ km/s/Mpc.  Orbit sampling frequencies 
are weighted by $b_{\rm Kep}$ to account
for the higher probability of orbits with larger impact
parameters.
Points are at random times, viewed from
random angles.  The open circles represent points just after
the first close pass; the filled circles are just after the 
second close pass. The galaxy masses are M$_{\rm gal} = 
5.8 \times 10^{11}$ M$_{\sun}$.
The orbits have zero initial orbital energy.}
\label{fig:or3}
\end{figure}

To this point, we have not used the measured velocity separation,
$\Delta V$, in our models.  In Fig.~\ref{fig:or3} we plot 
the expected $\Delta V$ vs. $t_{\rm pass}$ for the 
orbits of Fig.~\ref{fig:or1}.  At apogalacticon, the 
range of $\Delta V$ values is very tight around zero; before and after, 
the range is much broader.  The apogalacticon time changes somewhat
for different orbital energies or different galaxy masses.  

\begin{figure}
\centerline{\epsfysize=7in%
\epsffile{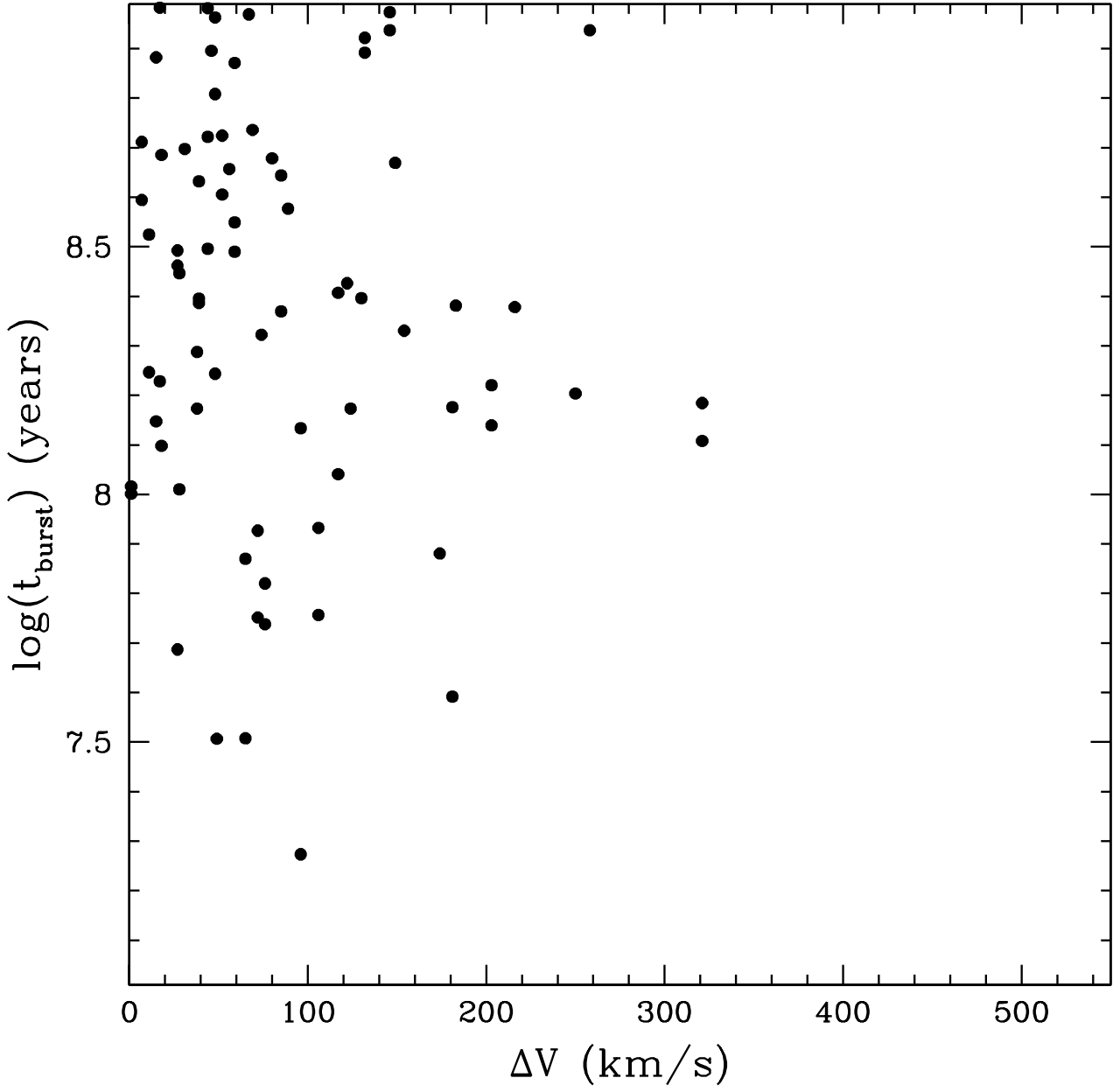}}
\caption{$\Delta V$ vs. log($t_{\rm pass}$) from the 
L99 model with instantaneous star
formation, Z=Z$_{\sun}$ and a Miller-Scalo IMF, for
${\rm t}_{\rm delay}=0$.  Only points with
$t_{\rm pass} < 10^9$~years and $\Delta V < 550$~km/s appear.}
\label{fig:or4}
\end{figure}

We plot $\Delta V$ vs. $\log(t_{\rm pass})$ for the data 
in Fig.~\ref{fig:or4},
using the L99 model of Fig.~\ref{fig:lh1}b, 
with $\alpha = 3.3$, Z=Z$_{\sun}$, continuous
star formation, and no delay (Fig.~\ref{fig:or4}a).  
Qualitatively, the distribution of the
points is similar to the predicted distribution (Fig.~\ref{fig:or3}):
the velocity separations are spread out 
for small $t_{\rm burst}$, especially 
for the HeI-emitting galaxies.  The velocity separations squeeze
in at $\log(t_{\rm pass,apo}) \sim 8.2$, and spread out 
again at later times.
However, the exact value of $t_{\rm pass,apo}$ is not determined by our
data: a delay, a different model,
or any physical process that can change the mapping from 
EW(H$\alpha$) to $t_{\rm pass}$ 
can change the $t_{\rm burst,apo}$ derived from the data.  

\subsection{After the Starburst}

If the galaxies with large H$\alpha$ equivalent width are
starbursting galaxies observed just
after a close pass, where are the post-starburst
galaxies?  Only one galaxy in our spectroscopic sample, NGC 7715,
has a classical ``E+A'' spectrum, with a very blue continuum, 
no emission whatsoever, and huge EW(H$\delta$)$ = -7.7$, 
but several galaxies show detectable H$\delta$ absorption, 
consistent with a fading, $\sim$$10^9$-year-old starburst,
when the galaxies should be at or past apogalacticon.
Thus, we might expect these post-burst candidates 
to have a range of $\Delta D$ values, determined largely
by the range of apogalacticon distances populating the sample
(see the range of $\Delta D$ for $t_{\rm pass} \approx 10^9$
in Fig.~\ref{fig:or1}).
However, if these galaxies are clumped near apogalacticon,
their $\Delta V$ distribution would be narrower than for 
pairs at other stages (see the smaller range of $\Delta V$
for $t_{\rm pass} \approx 10^9$ in Fig.~\ref{fig:or3}).

\begin{figure}
\centerline{\epsfysize=7in%
\epsffile{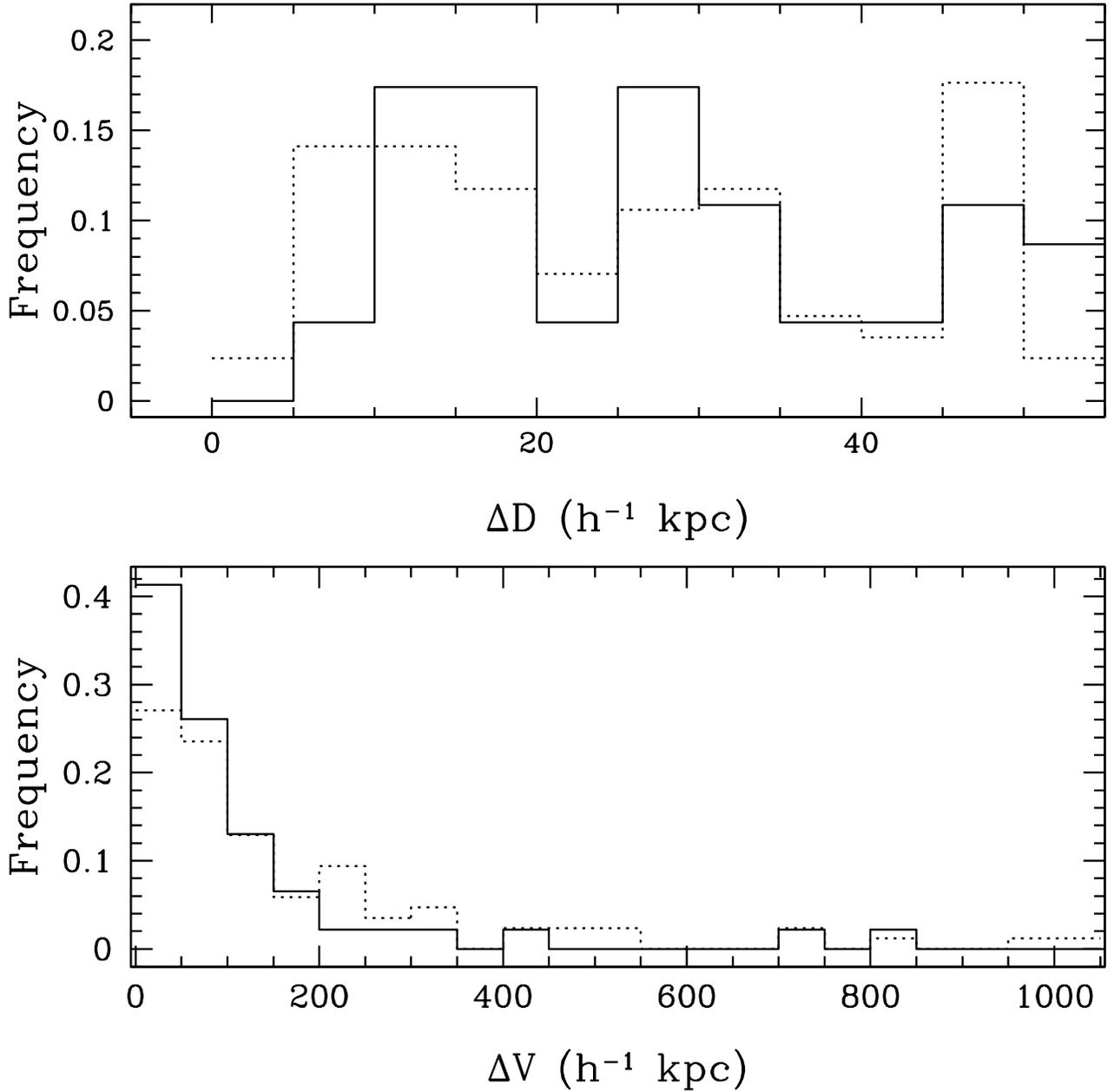}}
\caption{Balmer absorption: (a) $\Delta D$ distributions
of galaxies in low-density ($\rho_{2.5} \leq 2.2$)
environments  with EW(H$\delta) < -2$~\AA\ (solid line) and
EW(H$\delta) \geq 0$ (dotted line), and (b) $\Delta V$
distributions of the same subsamples.  Note that the sample
with EW(H$\delta) \geq 0$ includes both emission-line galaxies
and absorption galaxies.}
\label{fig:lh6}
\end{figure}

Fig.~\ref{fig:lh6} shows normalized
histograms of $\Delta D$ and
$\Delta V$ for the 46 galaxies in our low-density 
($\rho_{2.5} \leq 2.2$) sample with
EW(H$\delta$)~$< -2$~\AA\ (solid line) and for the 85 galaxies
in our low-density sample with EW(H$\delta$)~$\geq 0$ (dotted line).  
K-S tests show that the $\Delta D$ distributions are the same
at the 24\% level.  However, the $\Delta V$ distributions differ
at the 11.7\% level, and at the 3.7\% level if
restricted to $\rho_{2.5} \leq 2.0$.  The distribution for the points with
EW(H$\delta$))~$< -2$~\AA\ is clearly shifted towards 
smaller $\Delta V$, consistent with a population of galaxies
near apogalacticon.

NGC 7715, the true ``E+A'' galaxy, is paired with the
Wolf-Rayet galaxy NGC 7714 ($\Delta V = 49$~km/s). 
NGC 7714 shows a ring-like distortion
likely associated with a very close pass, or even a 
penetrating encounter (see, e.g., Papaderos \& Fricke 1998 for
a detailed discussion).

\subsection{Helium}

HeI is a recombination line which provides further constraints on and
consistency checks of the application of the L99 star-formation models.
The relative strength of HeI
can be estimated directly from the expected number of ionizing
photons given in the L99 models.  
The relative strength of HeI 
provides a good consistency check --- 
in some of the L99 models,
the predicted
ratio of HeI/H$\alpha$ matches the data well for the young
starbursts.

Under the assumption that the
nebulae surrounding the massive stars are 
optically thick to all photons in the Lyman and HeI continuum, and
that radiative transfer effects are negligible, 
the ratio HeI/H$\alpha$ can be estimated from the relevant photon 
populations and recombination coefficients, assuming 
n$_{\rm e} = 10^{4}{\rm\ cm}^{-3}$ and T$_{\rm e} = 10^4$~K, 
where n$_{\rm e}$ and T$_{\rm e}$ are the electron density and
temperature, respectively. 
The HeI($\lambda$5875)/H$\alpha$ ratio under these assumptions
is 0.28~N(He$^0$)/N(H$^0$), where N(He$^0$) and N(H$^0$) are the number 
of photons capable of ionizing HeI and HI, respectively (Osterbrock 1989).

\begin{figure}
\centerline{\epsfysize=7in%
\epsffile{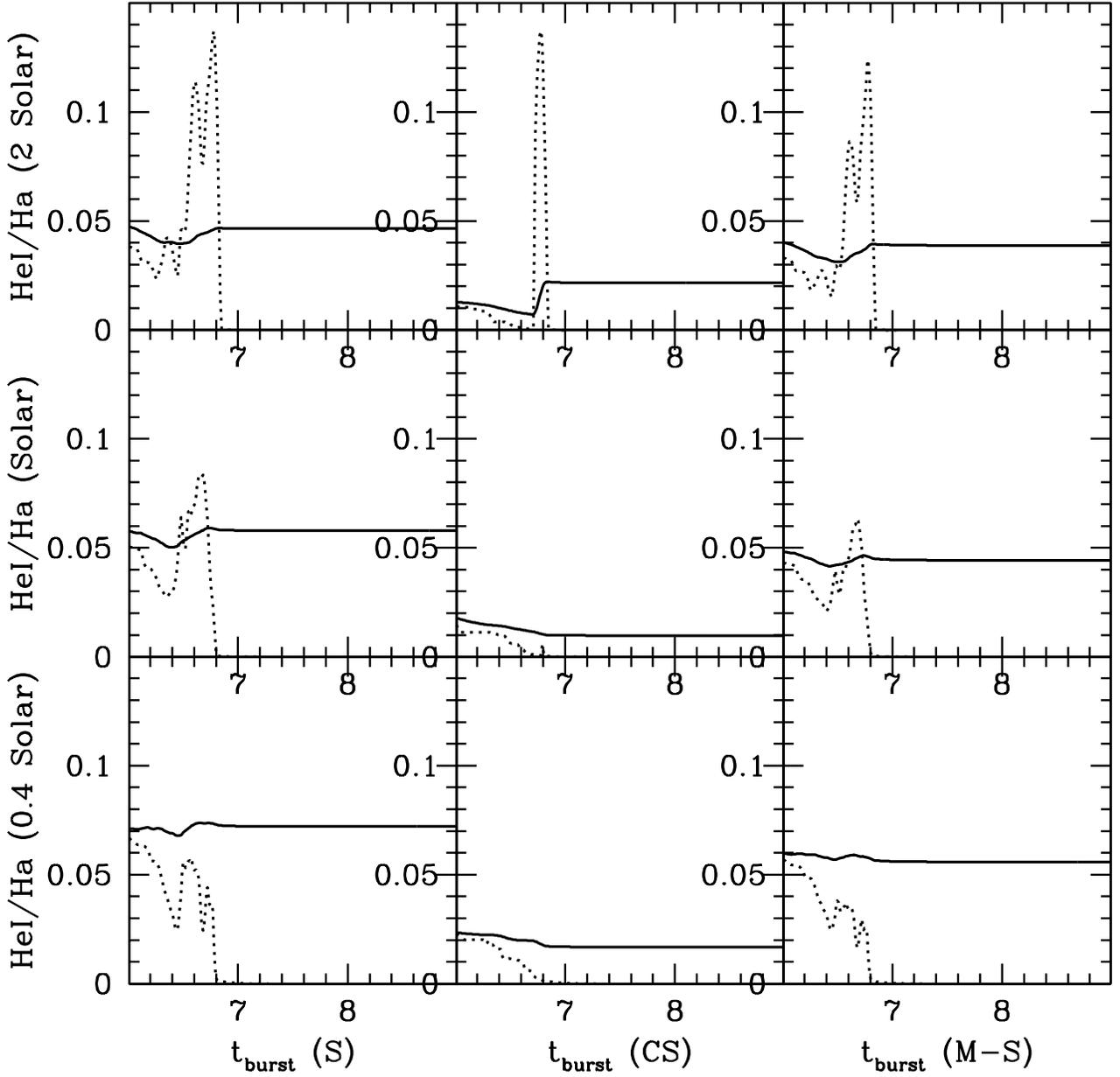}}
\caption{Predicted HeI$\lambda5876$/H$\alpha$ as a function of
time for the continuous (solid line) and instantaneous (dotted line)
star formation models of L99.
S, CS, and M-S refer to the Salpeter, Cutoff Salpeter, and
Miller-Scalo IMFs, respectively.}
\label{fig:he1}
\end{figure}

The L99 models predict both N(He$^0$) and N(H$^0$).  Fig.~\ref{fig:he1}
shows a grid of 0.28~N(He$^0$)/N(H$^0$), or the predicted
HeI/H$\alpha$, as a function of time from L99.
Each plot shows the continuous (solid line) and instantaneous (dotted line)
bursts.  The L99 models predict that HeI emission should 
disappear very rapidly after a burst, 
in $\lesssim 10^7$~years.  In continuous star-forming 
mode, HeI/H$\alpha$ stays roughly constant; the ratio deviates from
its continuous mode only when the star formation begins or ceases.

\begin{figure}
\plottwo{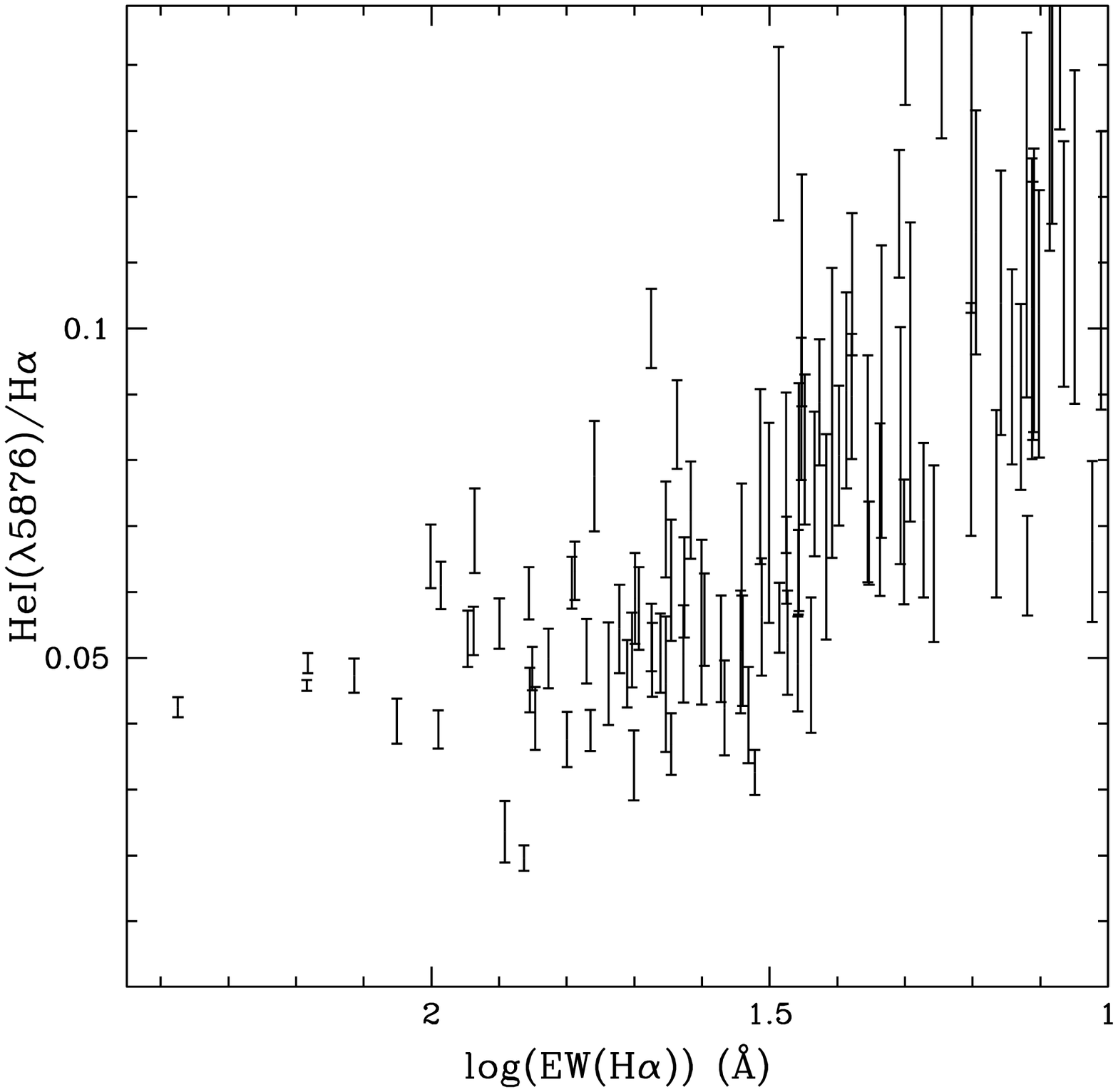}{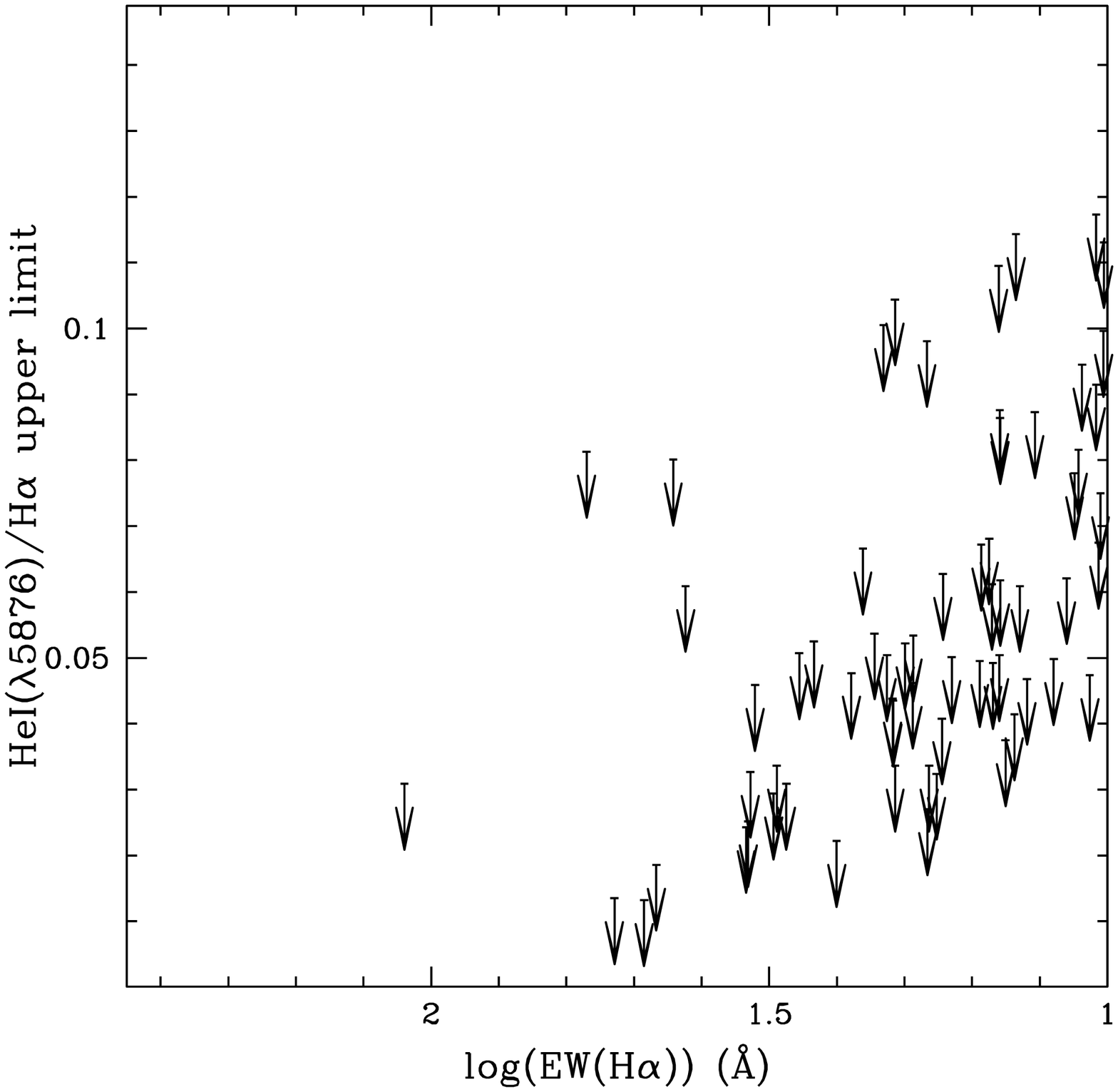}
\caption{Observed HeI/H$\alpha$ as a function of 
log(EW(H$\alpha$)) for EW(H$\alpha$)$ > 10$~\AA: 
(a) measured ratios, and (b) upper limits for 
the remaining galaxies.}
\label{fig:he2}
\end{figure}

Fig.~\ref{fig:he2}a shows HeI/H$\alpha$ vs. EW(H$\alpha$)
for pair galaxies in low-density environments ($\rho_{2.5} \leq 2.2$) in 
our sample with EW(H$\alpha$)$ > 10 $\AA.  We correct HeI/H$\alpha$
for 0.5~\AA\ of stellar absorption in Fig.~\ref{fig:he2}a 
(T. Heckman, private communication), although we neglect the 
effects of this small correction in Fig.~\ref{fig:he2}b (and
in Sec.~6).  The 
x axis is inverted;  younger starbursts are to
the left.  The galaxies with the largest 
EW(H$\alpha$) cluster around 
HeI/H$\alpha$~=~0.04~--~0.05, as predicted by several of the 
models in Fig.~\ref{fig:he1}.
Most notably, the best matches are for the 2 Z$_{\sun}$ 
Salpeter model, and the Miller-Scalo model with Z~=~1~--~2~Z$_{\sun}$.
The cutoff Salpeter model predicts HeI/H$\alpha$~$\leq 0.02$ for these
metallicities, which is below the measured values.
However, these results should be interpreted with caution.
L99 and Leitherer \& Heckman (1995) note that
N(He$^0$) is very sensitive to metallicity variations.
Their models are not chemically self-consistent --- the stellar
populations maintain the same metallicity throughout the
calculation.  

The scatter in HeI/H$\alpha$ increases for the older starbursts.
This increase could have many causes, including variations in 
metallicity, increased dust, or the inadequacy of the assumptions relating
HeI/H$\alpha$ to N(He$^0$)/N(H$^0$). 
Additionally, for small EW(H$\alpha$), non-Gaussian errors 
could contribute more significantly
(e.g. problems with the correction
for Balmer absorption which are not reflected
in the error bars).  However,
if these possibilities could be eliminated, the HeI/H$\alpha$
ratio could serve as an indicator of an abrupt change in the star 
formation rate.

In Fig.~\ref{fig:he2}b we show upper limits to HeI/H$\alpha$
for several galaxies
with H$\alpha$ emission but no detected HeI.  
In most cases, the limits are consistent
with the values in Fig.~\ref{fig:he2}a.  However, a few upper
limits, as well as a few HeI/H$\alpha$ ratios 
in Fig.~\ref{fig:he2}a, 
are quite small and should be investigated as
systems where star formation may have ceased within the last
$\sim$10$^7$ years.

In summary, the behavior of HeI generally agrees with the predictions
of the L99 models, especially some of the Miller-Scalo and 
Salpeter IMF models.  If the models predict the correct HeI/H$\alpha$, 
which depends on the detailed treatment of the most massive stars and 
is therefore currently uncertain,
the measured HeI/H$\alpha$ distribution 
can distinguish between the cutoff Salpeter and Miller-Scalo
IMFs, which both roughly match the $\Delta D$-$t_{\rm burst}$
distribution (see Sec.~7); the (uncertain) predictions of L99 favor the
Miller-Scalo IMF.  Additionally, outliers
in the  HeI/H$\alpha$ distribution may be galaxies which have undergone
a recent, abrupt change in the star formation rate.

\section{Discussion}

A large number of effects, in addition to model inaccuracy, can lead to 
scatter in the transformation from an H$\alpha$ equivalent width to a 
$t_{\rm burst}$, as well as to a $t_{\rm pass}$.  
Here, we estimate the scatter in the transformation and
mention the effects of old stellar populations and different 
galaxy potentials.

Paired galaxies, if they are truly interacting, 
have identical values of the time since a close pass, 
$t_{\rm pass}$.  If the delay between the pass and
the induced starburst is the same for the two galaxies, 
if neither have significant contamination at
H$\alpha$ from an old stellar population, and if both have
available gas, the galaxies in the
pair should have identical EW(H$\alpha$).  Kennicutt 
et al. (1987) report a correlation of EW(H$\alpha$)
in galaxies in the same pair, which they note is
similar to the ``Holmberg effect'' for concordance in galaxy 
colors (Holmberg 1958; Madore 1986).

\begin{figure}
\plottwo{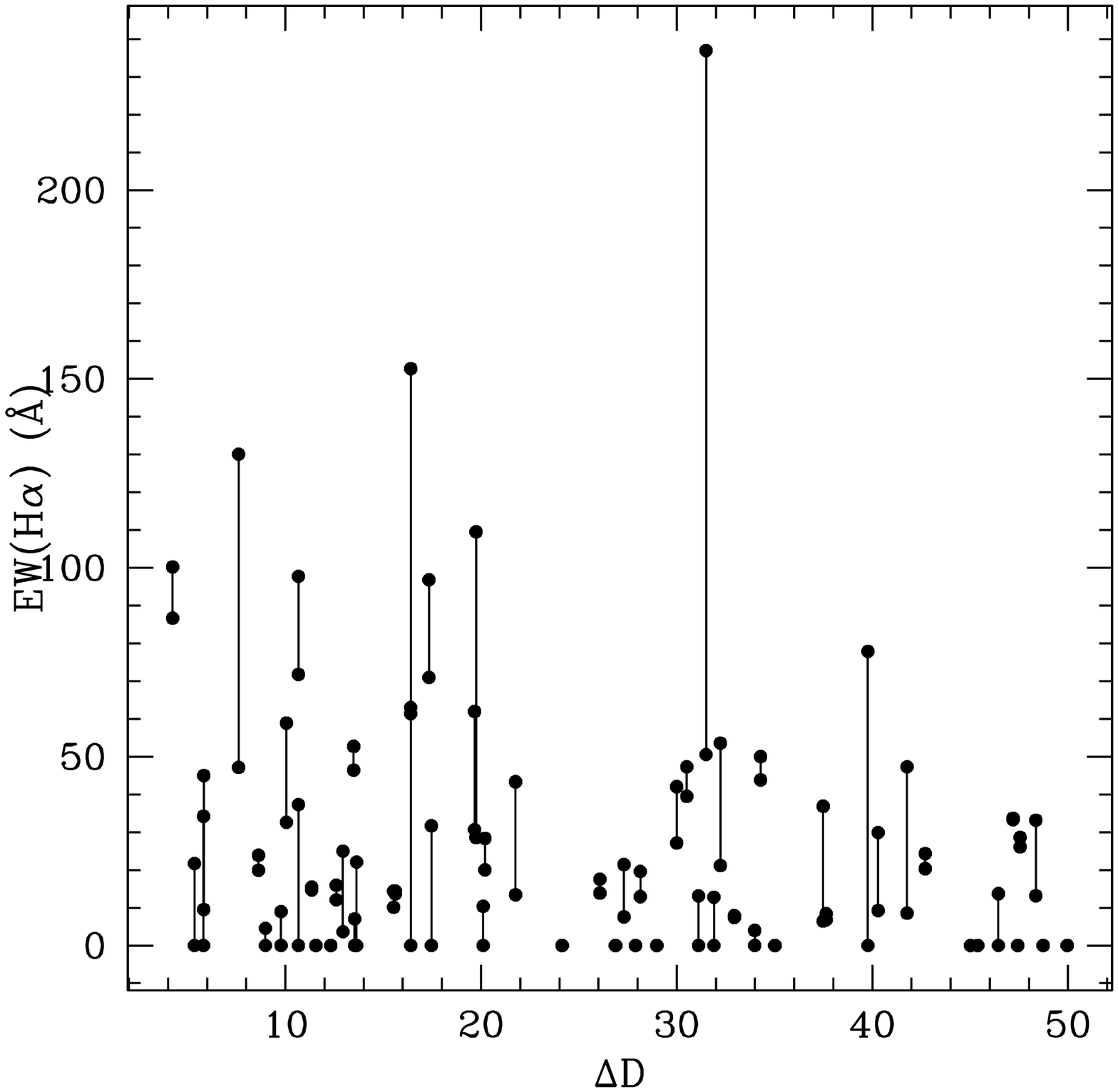}{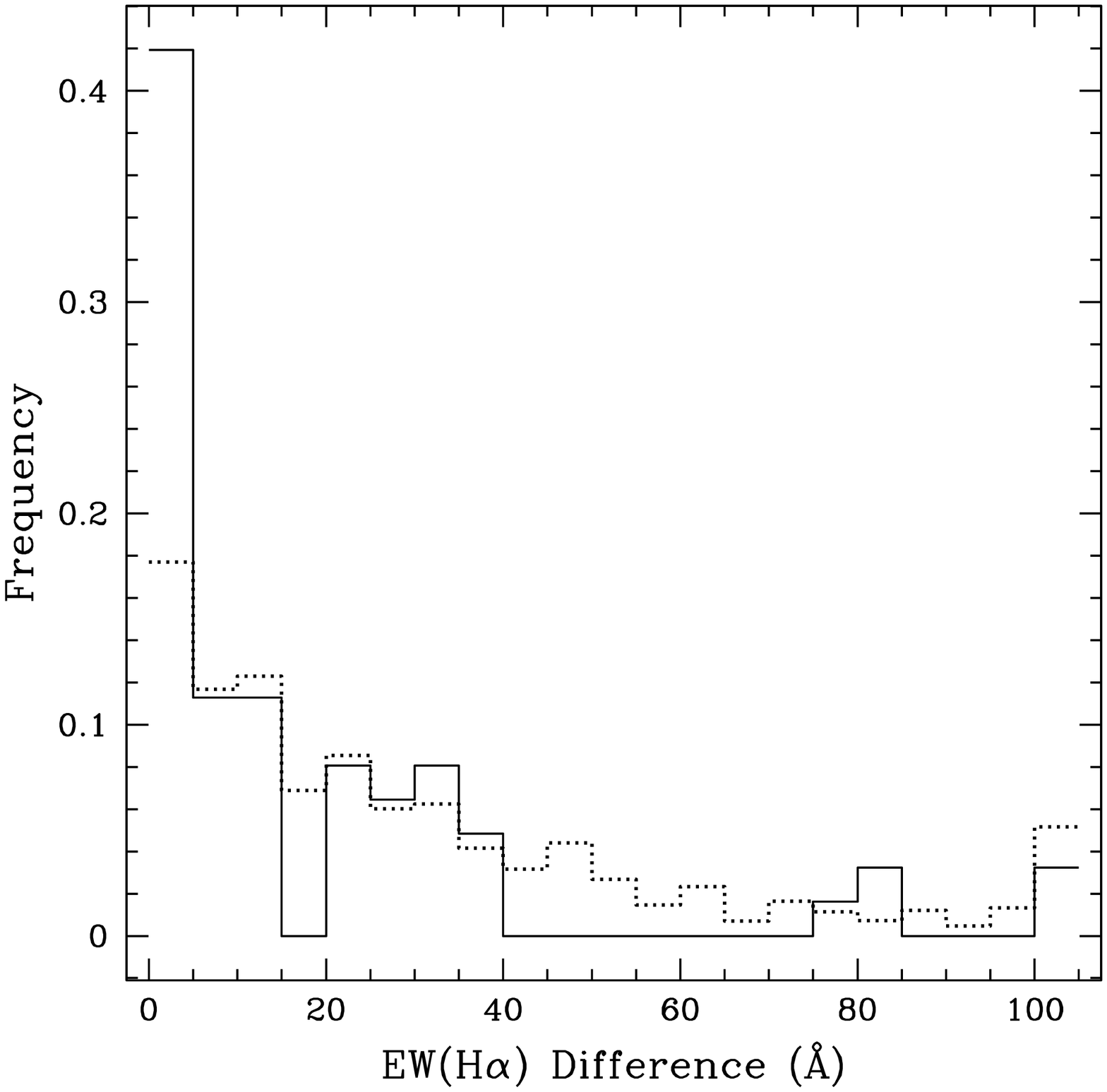}
\caption{Pairs in environments with $\rho_{2.5} \leq 2.2$: 
(a) $\Delta D$ vs. EW(H$\alpha$ with galaxies in the
same pair connected and (b) distribution of the EW(H$\alpha$)
scatter for the 66 pairs (solid line) and a ``random'' Monte
Carlo scatter (dotted line), computed by sampling from the same EW(H$\alpha$)
distribution.}
\label{fig:sc1}
\end{figure}

Some pairs in our sample have very similar EW(H$\alpha$);
others have widely discrepant values.  Fig.~\ref{fig:sc1}a shows
EW(H$\alpha$) as a function of $\Delta D$ with points in the
same pair connected; we restrict the sample to galaxies in
environments with $\rho_{2.5} \leq 2.2$ and include only
pairs (no n-tuples) for simplicity.
The median scatter in EW(H$\alpha$) for these 66 pairs is 
8.7 \AA.

We test the significance of similarity in EW(H$\alpha$) using
a Monte Carlo simulation in which we scramble the 
EW(H$\alpha$) values among the relevant pairs.  We compare the true
distribution of scatter in EW(H$\alpha$) to the Monte Carlo
version (from 500 scramblings) in Fig.~\ref{fig:sc1}b. The
true distribution includes a clear excess near zero; a K-S
test indicates that the distributions differ significantly,
at the 0.05\% level.  

We roughly test the possibility that the 
apparent concordance may arise from the $\Delta D$~--~EW(H$\alpha$)
correlation alone, since galaxies in pairs always have the
same $\Delta D$.  We apply the test to just the 27 pairs with 
$\Delta D < 20 {\rm\ h}^{-1}$~kpc and the 28 pairs
$\Delta D > 30 {\rm\ h}^{-1}$~kpc; the observed distributions
differ from the scrambled Monte Carlo distributions at
the 1.5\% and 2.5\% levels, respectively.  
We conclude, in agreement with Kennicutt et al. (1987), that
EW(H$\alpha$) is correlated for galaxies in the
same pairs; this concordance is probably not just an artifact
of the $\Delta D$~--~EW(H$\alpha$) correlation.

When comparing the points to the models in Sec.~7, the 
appropriate quantity to consider for each pair is the
largest EW(H$\alpha$) in the pair. 
Most factors serve only
to reduce EW(H$\alpha$) ``artificially''.
One exception is the star formation normally associated
with spiral galaxies, which will increase EW(H$\alpha$).
In our sample, some of these factors can be 
explored directly, by comparing galaxies with their
companions.

The most important factor affecting
$t_{\rm burst}$ may be the
presence of an old stellar population contributing to the 
continuum near H$\alpha$.  If it does, EW(H$\alpha$) 
depends somewhat on the burst strength; a correlation
could then arise between EW(H$\alpha$) and $\Delta D$ if 
pairs on small orbits have the strongest starbursts.
However, Barton et al. (1999b) combine photometry with
these spectra and use a two-population model for the
stellar content of the galaxies 
to show that the $\Delta D$~--~EW(H$\alpha$)
correlation is at least partly a result of the presence of
older bursts at larger $\Delta D$.  This conclusion
is relatively robust with respect to assumptions about
reddening, the old stellar population, and the IMF.
Using the L99 continuous Miller-Scalo model with Z=Z$_{\sun}$,
and some assumptions about the reddening and old stellar population,
they also show that the strengths of many of the bursts 
are probably large.  This analysis indicates that the old stellar population
would move the locus of the points in Fig.~\ref{fig:lh1}b by only
a few tenths of a unit along the y axis.

Photometry and rotation curves of the starbursting galaxies
in our sample will provide a further check of the MH96 
models (Barton et al. 1999c).
The MH96 models show that the response
of the gas to a close pass depends dramatically on the
mass distribution of a galaxy.  
Observations that extend beyond the nuclear region
(e.g. Kennicut et al. 1987) 
indicate that interactions can enhance the star 
formation in the disk, as well as the nucleus.  
Mihos (1997) models gas-rich, low surface brightness galaxies
and shows that the tidally-triggered star formation is
spatially extended. In the MH96 simulations of galaxies with
steeper potentials, the galaxies with deep central potentials and 
bulges respond only minimally to a close pass, 
while close passes trigger starbursts in 
galaxies with shallow central potentials (on prograde orbits).
Thus, if the picture from the simulations
is correct, we expect the galaxies in our sample with 
nuclear bursts of star formation to have no bulges
and shallow inner rotation curves; galaxies with extended
star formation should have even shallower rotation curves.
Longslit, major axis rotation curves taken at the MMT
(Barton et al. 1999c) of some of the emission-line
galaxies in our sample show qualitatively that the brightest
few starbursts have shallow rotation curves, with small velocity
widths, suggesting that their central potentials are shallow
and their total masses are small.  

Orbital inclination 
also affects the response of galaxies
to tidal interaction.  Toomre \& Toomre (1972) initially 
realized that in galaxy collisions, distortion is induced by 
resonant interactions.  Thus, the differences in response
between prograde and retrograde galaxies are dramatic --- 
retrograde galaxies show almost no tidal or kinematic 
distortions.  However, Keel (1993) reports no detectable
differences in EW(H$\alpha$)
between statistical samples of 27
mostly prograde and 21 mostly retrograde galaxies in
pairs.  MH96 report that retrograde encounters can trigger star
formation in simulations, although they do not produce dramatic 
tidal tails. Preliminary results from our sample indicate that prograde and
retrograde encounters have similar emission-line
characteristics, although the prograde encounters appear to
produce different tidal features.

We plan to discuss these issues in more detail
by incorporating B and R
images and rotation curves into the analysis (Barton et al. 1999c).
We will include a detailed analysis
of the surface brightness profiles and colors of the galaxies,
a measure of the steepness of the rotation curves, and a 
comparison of the prograde and retrograde encounters in 
our sample.

\section{Conclusion}

We analyze 502 new optical spectra of galaxies in pairs and n-tuples
in the original CfA2 redshift catalog, selected
to have projected separations $\Delta D \leq 50$h$^{-1}$~kpc, 
velocity separations $\Delta V \leq 1000$~km/s, no isolation
criteria, and redshifts $cz \geq 2300$~km/s.
The sample includes the complete set of pairs in the original
CfA2South survey and 45\% of all of the 1115 known galaxies in 
pairs and n-tuples in the region with m$_{\rm Zw} \leq 15.5$, 
included in the refined version of the CfA2 
survey (Falco et al. 1999).  

150 galaxies have significant emission in the four main
diagnostic lines: H$\alpha$, H$\beta$, [OIII]($\lambda5007$) and
[NII]($\lambda6584$).  Starbursts and HII-region galaxies are
common among these galaxies, but there is
no evidence for the excess of AGNs observed among
luminous infrared galaxies.  
Only 13\% of the emitting galaxies
in our sample have emission-line
ratios consistent with Seyferts or LINERS, compared with 41\% of
the LIG sample of Veilleux et al. (1995).  Conversely, the pair
sample contains a population of ``extreme starbursting galaxies''
absent from the LIGs of Veilleux et al. (1995).  

The equivalent widths of H$\alpha$ and other emission
lines depend strongly on pair $\Delta D$ and
$\Delta V$ (e.g., Kennicutt et al. 1987).  
These (anti-)correlations are not merely a result of large-scale
environmental effects that we detect; nor 
does the correlation result from other sample biases.  If we
assume EW(H$\alpha$) for an isolated burst of star formation
decreases monotonically with time, as starburst models like L99
suggest, and that this change in age dominates the
$\Delta D$~--~EW(H$\alpha$) correlation (Barton et al. 1999b),
then the correlation shows that galaxies in pairs
with larger separations tend to have older starbursts.

By matching the dynamical timescale to the 
burst timescale (L99), we show that the anti-correlation
between $\Delta D$ and EW(H$\alpha$) can arise from 
bursts of star formation initiated just after perigalacticon,  because
EW(H$\alpha$) decreases with time as the pair separation
increases.  
Our data show remarkable consistency with this picture:

\begin{enumerate}

\item EW(H$\alpha$) is highly correlated for galaxies in the same pair,
consistent with the fact that local interactions determine
the star-forming properties of galaxies in our sample (Kennicutt 
et al. 1987; for colors, see Holmberg 1958; Madore 1986).

\item The data agree with the expected distribution
of a population of orbiting 
galaxies in the $\Delta D$~--~$t_{\rm burst}$ plane 
for a Miller-Scalo or a cutoff (M$ \leq 30\ $M$_{\sun}$) Salpeter 
initial mass function, assuming that the contribution to
the continuum at H$\alpha$ from a very old stellar population is
minimal compared to the contribution from the most recent burst. 
The L99 Miller-Scalo model also predicts the HeI-to-H$\alpha$
ratios observed in our sample.

\item The data agree qualitatively
with the expected distribution
of galaxies in the $\Delta V$~--~$t_{\rm burst}$ plane.

\item We record a population of galaxies with Balmer absorption
that are candidate post-burst systems;  their
$\Delta V$ distribution is narrower than the $\Delta V$ distribution
of the other pairs, consistent with a population of galaxies
at apogalacticon, $\sim$$5\times 10^8$~--~$10^9$ years
after a close pass.

\end{enumerate}

The remarkable agreement between the data and many aspects of
the model shows that the
dynamical timescales of the pairs match the burst timescales
inferred from EW(H$\alpha$). 
The model is compatible with bursts of star formation 
induced by close passes
as long as the bursts last longer
than $\sim$10$^{8}$ years and the delays between the close passes and
the initiations of the bursts are less than a few$\times 10^7$ years.
N-body/SPH simulations (MH96) provide a physical basis for
this picture, predicting some starbursting galaxies in pairs; the simulations  
show that close interactions between galaxies 
can initiate bursts of star formation before the final merger if the galaxies have
shallow central potentials.  These starbursts should be 
visible while the galaxies are a resolvable pair.

The picture we develop here,
in conjunction with a large sample of spectroscopic and
photometric data, along with rotation curves, has great
potential for constraining the IMF and the duration
of tidally-triggered starbursts,
the true fraction of interacting galaxies in pairs, 
the orbits, and the dynamical timescales of the interactions.

\begin{acknowledgements}

We thank William Keel, Timothy Heckman, Claus Leitherer, and 
Robert Kennicutt for thoughtful comments which contributed 
substantially to this paper.
We thank Perry Berlind and Mike Calkins for assistance with observations
and Susan Tokarz for assistance with data reduction.  
We thank Norman Grogin for the use of his smoothed survey density
estimator, and for useful suggestions.  We are also very grateful to 
C. Leitherer and T. Heckman, as well as the other authors
of Leitherer et al. (1999), for making their valuable
starburst models readily available.  EJB received support from a 
Harvard Merit Fellowship.
This research is supported in part by the Smithsonian Institution.
This research has made use of the 
NASA/IPAC Extragalactic Database (NED)
which is operated by the Jet Propulsion Laboratory, California Institute
of Technology, under contract with the National Aeronautics and Space
Administration.

\end{acknowledgements}

\begin{deluxetable}{lrrl}
\tablenum{1}
\tablewidth{0pc}
\tablecolumns{4}
\tablecaption{K-S Tests of North/South Sample Differences}
\tablehead{ \colhead{Distribution} & \colhead{N$_{\rm gal}$ North} & \colhead{N$_{\rm gal}$ South} & \colhead{K-S Probability} }
\startdata
EW(H$\alpha$), all galaxies                          & 192 & 310 & $1.3 \times 10^{-2}$ \nl
EW(H$\alpha$), H$\alpha$-emitting galaxies           & 103 & 128 & $3.8 \times 10^{-1}$  \nl
$\Delta D$, all galaxies                             & 192 & 310 & $1.2 \times 10^{-2}$  \nl
$\Delta D$, H$\alpha$-emitting galaxies              & 103 & 128 & $2.0 \times 10^{-1}$  \nl
$\Delta D$, non-H$\alpha$-emitting galaxies          &  89 & 182 & $1.5 \times 10^{-1}$  \nl
$\Delta V$, all galaxies                             & 192 & 310 & $3.1 \times 10^{-2}$  \nl
$\Delta V$, H$\alpha$-emitting galaxies              & 103 & 128 & $2.4 \times 10^{-2}$  \nl
$\Delta V$, H$\alpha$-emitting galaxies, low density & 56  &  50 & $5.2 \times 10^{-3}$  \nl
\tablecomments{K-S tests of the differences between the North and South
distributions of EW(H$\alpha$), $\Delta D$, and $\Delta V$ for the full sample
and for various subsamples.}
\enddata
\label{tab:se}
\end{deluxetable}

\begin{deluxetable}{crrrrrr}
\tablenum{2}
\tablewidth{0pc}
\tablecolumns{7}
\tablecaption{Spearman Rank Probabilities of No Correlation}
\tablehead{ 
 & \multicolumn{2}{c}{All Densities} & \multicolumn{2}{c}{$\rho_{2.5} \leq 2.2$} & \multicolumn{2}{c}{$\rho_{2.5} \leq 2.2$, With H$\alpha$} \\
 & \multicolumn{2}{c}{(502 galaxies)} & \multicolumn{2}{c}{(162 galaxies)} & \multicolumn{2}{c}{(106 galaxies)}  \\ 
\colhead{Line} & \colhead{$\Delta D$~--~EW} & \colhead{$\Delta V$~--~EW} & \colhead{$\Delta D$~--~EW}  & \colhead{$\Delta V$~--~EW} & \colhead{$\Delta D$~--~EW}  & \colhead{$\Delta V$~--~EW} }
\startdata
H$\alpha$            & $5.2\times 10 ^{-3}$ & $6.3\times 10^{-5}$ & $1.7\times 10 ^{-2}$ & $1.9\times 10^{-3}$  & $7.0 \times 10^{-2}$ & $2.8 \times 10^{-1}$ \nl
H$\beta$             & $1.4\times 10 ^{-2}$ & $4.2\times 10^{-4}$ & $4.6\times 10 ^{-2}$ & $5.4\times 10^{-3}$  & $1.1 \times 10^{-1}$ & $1.5 \times 10^{-1}$ \nl
HeI($\lambda5876$)   & $3.1\times 10 ^{-2}$ & $9.7\times 10^{-3}$ & $1.5\times 10 ^{-2}$ & $1.4\times 10^{-2}$  & $1.3 \times 10^{-2}$ & $1.6 \times 10^{-1}$ \nl
OII ($\lambda3727$)  & $6.7\times 10 ^{-3}$ & $5.8\times 10^{-5}$ & $1.5\times 10 ^{-1}$ & $4.1\times 10^{-3}$  & $3.7 \times 10^{-1}$ & $9.1 \times 10^{-2}$ \nl
OIII ($\lambda5007$) & $1.6\times 10 ^{-3}$ & $5.7\times 10^{-6}$ & $3.6\times 10 ^{-2}$ & $1.9\times 10^{-3}$  & $9.4 \times 10^{-2}$ & $8.5 \times 10^{-2}$ \nl
\tablecomments{P$_{\rm SR}$ for possible correlations between orbit parameters
($\Delta D$,$\Delta V$) and equivalent widths (EWs) for various lines, including all the galaxies, 
all the galaxies in environments with $\rho_{2.5} \leq 2.2$, or only the galaxies with 
significant H$\alpha$ emission in environments with $\rho_{2.5} \leq 2.2$.}
\enddata
\label{tab:ol1}
\end{deluxetable}

\clearpage

\clearpage

\end{document}